\documentclass[aps,groupedaddress,showpacs,floatfix]{revtex4}
\usepackage{amssymb}
\usepackage{amsmath}
\usepackage{graphicx}
\DeclareMathOperator{\sgn}{sgn}
\DeclareMathOperator{\Ai}{Ai}
\begin{document}

\title{Replica Bethe ansatz derivation of the Tracy-Widom distribution 
of the free energy fluctuations in one-dimensional 
directed polymers}

\author{Victor Dotsenko$^{\, a,b}$ }

\affiliation{$^a$LPTMC, Universit\'e Paris VI, 75252 Paris, France}

\affiliation{$^b$L.D.\ Landau Institute for Theoretical Physics,
   119334 Moscow, Russia}

\date{\today}

\begin{abstract}

 The distribution function of the free energy fluctuations 
in one-dimensional directed polymers with 
$\delta$-correlated random potential is studied by mapping the
replicated problem to the $N$-particle quantum boson system with attractive
interactions.  We find the full set of eigenfunctions and eigenvalues
of this many-body system and perform the summation over the entire spectrum of excited
states. It is shown that in the thermodynamic limit the problem is reduced 
to the Fredholm determinant with
the Airy kernel yielding the universal Tracy-Widom distribution,
which is known to describe the statistical properties
of the Gaussian unitary ensemble as well as many other statistical systems.

\end{abstract}

\pacs{
      05.20.-y  
      75.10.Nr  
      74.25.Qt  
      61.41.+e  
     }

\maketitle

\medskip

\begin{center}
 
{\bf \large Contents}

\end{center}

\begin{tabbing}
 AAAAAA \= {\bf Appendix C: Wave functions of quantum bosons with attractive interactions} \=
 . . . . \=  13\kill

\> {\bf I. Introduction}                                               \> . . . .    \> 1  \\
\> {\bf II. Mapping to quantum bosons}                                 \> . . . .    \> 3  \\
\> {\bf III. Replica partition function and the free energy distribution function} \> . . . .    \> 5 \\
\> {\bf IV. Conclusions}                                                        \> . . . .  \> 7 \\
\\
\> {\bf Appendix A: Quantum bosons with repulsive interactions}                     \>  \>  \\
\> \hspace{5mm}    1. Eigenfunctions                                            \> . . . . \> 8 \\
\> \hspace{5mm}    2. Othonormality                                             \> . . . . \> 9\\
\> {\bf Appendix B: Quantum bosons with attractive interactions}                    \>  \>  \\
\> \hspace{5mm}    1. Ground state                                              \> . . . . \> 10  \\
\> \hspace{5mm}    2. Eigenfunctions                                            \> . . . . \> 12  \\
\> \hspace{5mm}    3. Othonormality                                             \> . . . . \> 14  \\
\> \hspace{5mm}    4. Propagator                                                \> . . . . \> 18  \\
\> {\bf Appendix C: Fredholm determinant with the Airy kernel}                   \>         \>     \\
\>  \hspace{23mm}  {\bf  and the Tracy-Widom distribution}                          \> . . . . \> 20 \\
\end{tabbing}

\medskip

\section{Introduction}

Directed polymers in a quenched random potential have 
been the subject of intense investigations during the past two
decades (see e.g. \cite{hh_zhang_95}). 
Diverse physical systems such as domain walls in magnetic films
\cite{lemerle_98}, vortices in superconductors \cite{blatter_94}, wetting
fronts on planar systems \cite{wilkinson_83}, or Burgers turbulence
\cite{burgers_74} can be mapped to this model, which exhibits numerous
non-trivial features deriving from the interplay between elasticity and
disorder.
In the most simple one-dimensional case
we deal with an elastic string  directed along the $\tau$-axis 
within an interval $[0,L]$. Randomness enters the problem 
through a disorder potential $V[\phi(\tau),\tau]$, which competes against 
the elastic energy.  The problem is defined by the Hamiltonian
\begin{equation}
   \label{bas1-1}
   H[\phi(\tau), V] = \int_{0}^{L} d\tau
   \Bigl\{\frac{1}{2} \bigl[\partial_\tau \phi(\tau)\bigr]^2 
   + V[\phi(\tau),\tau]\Bigr\};
\end{equation}
where in the simplest case the disorder potential $V[\phi,\tau]$ 
is Gaussian distributed with a zero mean $\overline{V(\phi,\tau)}=0$ 
and the $\delta$-correlations: 
\begin{equation}
   \label{bas1-2}
{\overline{V(\phi,\tau)V(\phi',\tau')}} = u \delta(\tau-\tau') \delta(\phi-\phi')
\end{equation}
Here the parameter $u$ describes the strength of the disorder.
Historically, the problem of central interest was the scaling behavior of the 
polymer mean squared displacement  which in the thermodynamic limit
($L \to \infty$) is believed to have a universal scaling form
$\overline{\langle\phi^{2}\rangle}(L) \propto L^{2\zeta} $
(where $\langle \dots \rangle$ and $\overline{(\dots)}$ denote 
the thermal and the disorder averages), with $\zeta=2/3$, the so-called wandering exponent. 
More general problem for all directed polymer systems of the type, Eq.(\ref{bas1-1}),
is the statistical properties  of their free energy fluctuations. 
Besides the usual extensive (linear in $L$) self-averaging part $f_{0} L$ 
(where $f_{0}$ is the linear free energy density),
the total free energy $F$ of such systems contains disorder dependent 
fluctuating contribution $\tilde{F}$, which is characterized by
non-trivial scaling in $L$. It is generally believed  that in the limit of large $L$ 
the typical value of the free energy fluctuations scales with $L$ as
$\tilde{F}  \propto L^{1/3}$ (see e.g. \cite{hhf_85,numer1,numer2,kardar_87}).
In other words, in the limit of large $L$ the total (random) free energy of the system 
can be represented as
\begin{equation}
\label{bas1-5}
F \; = \; f_{0} L \; + \; c \, L^{1/3}\; f
\end{equation}
where $c$ is the parameter, which depends on the temperature and the strength of disorder, and 
$f$ is the random quantity which in the thermodynamic limit $L\to\infty$ 
is described by a non-trivial universal
distribution function $P_{*}(f)$. The derivation of this function for the system with $\delta$-correlated
random potential, Eqs.(\ref{bas1-1})-(\ref{bas1-2}) is the central issue of the present work.

 For the string with the zero boundary conditions
at $\tau=0$ and at $\tau=L$ the partition function of a given sample is
\begin{equation}
\label{bas1-6}
   Z[V] = \int_{\phi(0)=0}^{\phi(L)=0} 
              {\cal D} [\phi(\tau)]  \;  \mbox{\Large e}^{-\beta H[\phi,V]}
\end{equation}
where $\beta$ denotes the inverse temperature. On the other hand,
the partition function is related to the total free energy $F[V]$ via
\begin{equation}
\label{bas1-7}
Z[V] = \exp( -\beta  F[V])
\end{equation}
The free energy $F[V]$  is defined for a specific 
realization of the random potential $V$ and thus represent a random variable. 
Taking the $N$-th power of both sides of this relation 
and performing the averaging over the random potential $V$ we obtain
\begin{equation}
\label{bas1-8}
\overline{Z^{N}[V]} \equiv Z[N,L] = \overline{\exp( -\beta N F[V]) }
\end{equation}
where the quantity in the lhs of the above equation is called the replica partition function.
Substituting here $F = f_{0} L +  c \, L^{1/3}\; f$, 
 and redefining 
\begin{equation}
\label{bas1-10}
Z[N,L] = \tilde{Z}[N,L] \; \mbox{\Large e}^{-\beta N f_{0} L} 
\end{equation}
we get
\begin{equation}
\label{bas1-11}
\tilde{Z}[N,L] = \overline{\exp( -\lambda N f) }
\end{equation}
where $\lambda = \beta c L^{1/3}$.
The averaging in the rhs of the above equation can be represented in terms of the 
distribution function $P_{L}(f)$ (which depends on the system size $L$). 
In this way we arrive to the following general relation 
between the replica partition function $\tilde{Z}[N,L]$ and the distribution function 
of the free energy fluctuations $P_{L}(f)$:
\begin{equation}
\label{bas1-12}
   \tilde{Z}[N,L] \; =\;
           \int_{-\infty}^{+\infty} df \, P_{L} (f) \;  
           \mbox{\Large e}^{ -\lambda N   \, f}
\end{equation}
Of course, the most interesting object is the thermodynamic limit 
distribution function $P_{*}(f) = \lim_{L\to\infty} P_{L} (f)$ which is expected to be 
the universal quantity. The above equation is the bilateral Laplace transform of  the function $P_{L}(f)$,
and at least formally it allows to restore this function via inverse Laplace transform 
 of the replica partition function $\tilde{Z}[N,L]$. In order to do so one has to compute 
$\tilde{Z}[N,L]$ for an  arbitrary 
integer $N$ and then perform  analytical continuation of this function 
from integer to arbitrary complex values of $N$. 
In Kardar's original solution \cite{kardar_87}, after
mapping the replicated problem to interacting quantum bosons, one arrives at
the replica partition function for positive integer parameters $N > 1$.
Assuming a large $L \to \infty$ limit, one is tempted to approximate the
result by the ground state contribution only, as for any $N > 1$ the
contributions of excited states are exponentially small for $L\to\infty$.
However, in the analytic continuation for arbitrary complex $N$ the
contributions which are exponentially small at positive integer $N > 1$ can
become essential in the region $N\to 0$, which  defines the function $P(f)$
(in other word, the problem is that the two limits $L\to\infty$ and $N\to 0$ 
do not commute \cite{Medina_93,dirpoly}). As a consequence,
the resulting distribution function $P_{*}(f)$ 
exhibits severe pathologies such as the vanishing of its second 
moment, which assumes that the distribution function is not positively defined.
Nevertheless,  in terms of this approximation (assuming he universal scaling $L^{1/3}$ 
of the free energy fluctuations)
 one can derive the  left tail of the distribution function
$P_{*}(f)$ which is given by the asymptotics of the Airy function,
$P_{*}(f\to -\infty) \sim \exp\bigl(-\frac{2}{3}\big|f\big|^{3/2}\bigr)$ \cite{Zhang}.
For the first time the form of the {\it right tail} of this distribution function,
$P_{*}(f\to +\infty) \sim \exp\bigl[-(const) f^{3}\bigr]$,
has been derived in terms of the optimal fluctuation approach 
\cite{KK}, where it has been 
demonstrated that both asymptotics (left and right) of the 
function $P_{*}(f)$ are consistent with the Tracy-Widom (TW)
distribution \cite{Tracy-Widom}. Originally the TW distribution has been derived in the context 
of the statistical properties of the Gaussian Unitary Ensemble
while at present it is well established to 
describe the statistics of fluctuations in various random systems
\cite{PNG_Spohn,LIS,LCS,oriented_boiling,ballistic_decomposition,DP_johansson},
which were widely believed to belong to the same universality class as the
present model \cite{KK,Derrida1,Prahofer-Spohn}. 
Finally, quite recently, the one-point distribution of the solutions of the KPZ-equation 
(which is equivalent to the present model) has been derived for an arbitrary value 
of $L$, and it has been shown that in the limit $L \to \infty$ this distribution turns into 
the Tracy-Widom distribution \cite{KPZ-TW1,KPZ-TW2}.

In the  recent paper \cite{BAS} an attempt has been made to 
derive the free energy distribution function $P_{*}(f)$ via the calculation
of the replica partition function $Z[N,L]$ in terms of the 
Bethe-Ansatz solution for quantum bosons with attractive $\delta$-interactions
which involved the summation over the {\it entire spectrum} of
exited states. Unfortunately, the attempt has failed because 
on one hand, the calculations contained a kind of a hidden "uncontrolled  approximation",
and on the other hand, the analytic continuation of obtained $Z(N,L)$ 
was found  to be ambiguous.
Later on it turned out that it is possible to bypass the problem 
of the analytic continuation if instead of the
distribution function itself one would study its integral
representation
\begin{equation}
 \label{tw1}
W(x) \; = \; \int_{x}^{\infty} \; df \; P_{*}(f)  
\end{equation}
which gives the probability to find the fluctuation $f$ bigger that a given value $x$.
Formally the function $W(x)$ can be defined as follows:
\begin{equation}
 \label{tw2}
W(x) = \lim_{\lambda\to\infty} \sum_{N=0}^{\infty} \frac{(-1)^{N}}{N!} 
\exp(\lambda N x) \; \overline{\tilde{Z}^{N}} 
\; = \;  \lim_{\lambda\to\infty} \overline{\exp\bigl[-\exp\bigl(\lambda (x-f)\bigr)\bigr]} 
\; = \;  
\overline{\theta(f-x)}
\end{equation}
On the other hand, the probability function $W(x)$ 
can be computed in terms of the replica partition function
$\tilde{Z}[N,L]$ by summing over all replica integers
\begin{equation}
 \label{tw3}
W(x) \; = \; \lim_{\lambda\to\infty} \sum_{N=0}^{\infty} \frac{(-1)^{N}}{N!} 
\exp(\lambda N x) \; \tilde{Z}[N,L]
\end{equation}
In terms of this approach, quite recently it has been reported that
performing the summation over the entire spectrum of excited
states of $N$-particle quantum boson system  the above series, eq.(\ref{tw3}), 
can reduced to the Fredholm determinant with the  Airy kernel 
which is known to yield the  
Tracy-Widom distribution \cite{tw,LeDoussal}.

  Present paper provides the detailed derivation of this result. 
In particular, Appendices A and B contain detailed analysis of the 
properties of  $N$-particle wave functions  of quantum bosons 
with both repulsive and attractive $\delta$-interactions.
Besides, basing on the definition of the Fredholm determinant
with the Airy kernel, in Appendix C it is demonstrated how  
the explicit form of the Tracy-Widom distribution can be derived 
in terms of the solution of the Panlev\'e II differential equation.

\section{Mapping to quantum bosons}

Explicitly, the replica partition function, Eq.(\ref{bas1-8}), of the system described by
the Hamiltonian, Eq.(\ref{bas1-1}), is
\begin{equation}
\label{bas2-1}
   Z(N,L) = \prod_{a=1}^{N} \int_{\phi_{a}(0)=0}^{\phi_{a}(L)=0} 
   {\cal D} \phi_{a}(\tau) \;
   \overline{\exp\Biggl[-\beta \int_{0}^{L} d\tau \sum_{a=1}^{N}
   \bigl\{\frac{1}{2} \bigl[\partial_\tau \phi_{a}(\tau)\bigr]^2 
   + V[\phi_{a}(\tau),\tau]\bigr\}\Biggr] }
\end{equation}
Since  the random potential $V[\phi,\tau]$ has the Gaussian 
distribution the disorder averaging $\overline{(...)}$ in the above equation 
is very simple:
\begin{equation}
\label{bas2-2}
\overline{\exp\Biggl[-\beta \int_{0}^{L} d\tau \sum_{a=1}^{N}
     V[\phi_{a}(\tau),\tau]\Biggr] } \; = \; 
\exp\Biggl[\frac{\beta^{2}}{2} \int\int_{0}^{L} d\tau d\tau' \sum_{a,b=1}^{N}
     \overline{V[\phi_{a}(\tau),\tau] V[\phi_{b}(\tau'),\tau']}\Biggr]
\end{equation}
Using Eq.(\ref{bas1-2}) we find:
\begin{equation}
\label{bas2-3}
   Z(N,L) = \prod_{a=1}^{N} \int_{\phi_{a}(0)=0}^{\phi_{a}(L)=0} 
   {\cal D} \phi_{a}(\tau) \;
   \exp\Biggl[-\frac{1}{2}\beta \int_{0}^{L} d\tau 
   \Bigl\{\sum_{a=1}^{N} \bigl[\partial_\tau \phi_{a}(\tau)\bigr]^2 
   -\beta u \sum_{a,b=1}^{N} \delta\bigl[\phi_{a}(\tau)-\phi_{b}(\tau)\bigr]\Bigr\}\Biggr] 
\end{equation}
It should be noted that the second term in the exponential 
of the above equation contain formally divergent contributions proportional
to $\delta(0)$ (due to the terms with $a=b$). In fact, this is just an indication
that the {\it continuous} model, Eqs.(\ref{bas1-1})-(\ref{bas1-2}) is ill defined
as short distances and requires proper lattice regularization. Of course, the 
corresponding lattice model would contain no divergences, and the terms with
$a=b$ in the exponential of the corresponding replica partition function would produce
irrelevant constant $\frac{1}{2}L\beta^2 u N \delta(0)$ (where the lattice
version of $\delta(0)$ has a finite value). Since the  lattice 
regularization has no impact on the continuous long distance properties 
of the considered system this term will be just omitted in our further study.

Introducing the $N$-component scalar field replica Hamiltonian
\begin{equation}
\label{bas2-4}
   H_{N}[{\boldsymbol \phi}] =  
   \frac{1}{2} \int_{0}^{L} d\tau \Biggl(
   \sum_{a=1}^{N} \bigl[\partial_\tau\phi_{a}(\tau)\bigr]^2 
   - \beta u \sum_{a\not= b}^{N} 
   \delta\bigl[\phi_{a}(\tau)-\phi_{b}(\tau)\bigr] \Biggr)
\end{equation}
for the replica partition function, Eq.(\ref{bas2-3}), 
we obtain the standard expression
\begin{equation}
   \label{bas2-5}
   Z(N,L) = \prod_{a=1}^{N} \int_{\phi_{a}(0)=0}^{\phi_{a}(L)=0} 
   {\cal D} \phi_{a}(\tau) \;
   \mbox{\Large e}^{-\beta H_{N}[{\boldsymbol \phi}] }
\end{equation}
where ${\boldsymbol \phi} \equiv \{\phi_{1},\dots, \phi_{N}\}$.
According to the above definition this partition function describe the statistics
of $N$ $\delta$-interacting (attracting) trajectories $\phi_{a}(\tau)$ all starting 
(at $\tau=0$) and ending (at $\tau=L$) at zero: $\phi_{a}(0) = \phi_{a}(L) = 0$

In order to map the problem to one-dimensional quantum bosons, 
 let us introduce more general object
\begin{equation}
   \label{bas2-6}
   \Psi({\bf x}; t) = 
\prod_{a=1}^{N} \int_{\phi_a(0)=0}^{\phi_a(t)=x_a} {\cal D} \phi_a(\tau)
  \;  \mbox{\Large e}^{-\beta H_{N} [{\boldsymbol \phi}]}
\end{equation}
which describes $N$ trajectories $\phi_{a}(\tau)$ all starting at zero ($\phi_{a}(0) = 0$),
but ending at $\tau = t$ in arbitrary given points $\{x_{1}, ..., x_{N}\}$.
One can easily show that instead of using the path integral, $\Psi({\bf x}; t)$
may be obtained as the solution of the  linear differential equation
\begin{equation}
   \label{bas2-7}
\partial_t \Psi({\bf x}; t) \; = \;
\frac{1}{2\beta}\sum_{a=1}^{N}\partial_{x_a}^2 \Psi({\bf x}; t)
  \; + \; \frac{1}{2}\beta^2 u \sum_{a\not=b}^{N} \delta(x_a-x_b) \Psi({\bf x}; t)
\end{equation}
with the initial condition 
\begin{equation}
   \label{bas2-8}
\Psi({\bf x}; 0) = \Pi_{a=1}^{N} \delta(x_a)
\end{equation}
One can easily see that Eq.(\ref{bas2-7}) is the imaginary-time
Schr\"odinger equation
\begin{equation}
   \label{bas2-9}
-\partial_t \Psi({\bf x}; t) = \hat{H} \Psi({\bf x}; t)
\end{equation}
with the Hamiltonian
\begin{equation}
   \label{bas2-10}
   \hat{H} = 
    -\frac{1}{2\beta}\sum_{a=1}^{N}\partial_{x_a}^2 
   -\frac{1}{2}\beta^2 u \sum_{a\not=b}^{N} \delta(x_a-x_b) 
\end{equation}
which describes  $N$ bose-particles of mass $\beta $ interacting via
the {\it attractive} two-body potential $-\beta^2 u \delta(x)$. 
The original replica partition function, Eq.(\ref{bas2-5}), then is obtained via a particular
choice of the final-point coordinates,
\begin{equation}
   \label{bas2-11}
   Z(N,L) = \Psi({\bf 0};L).
\end{equation}


The eigenfunction of $N$-particle system of {\it repulsive} bosons ($u < 0$),
 Eq.(\ref{bas2-10}), have been derived by Lieb and Liniger in 1963 \cite{Lieb-Liniger}
(for details see Appendix A, as well as Refs. \cite{bogolubov, gaudin}). 
The spectrum and some properties of the eigenfunctions
for attractive ($u > 0$) one-dimensional quantum boson system have been derived by McGuire
\cite{McGuire} and by Yang \cite{Yang} (see also Ref.\ \cite{Takahashi,Calabrese}).
Detailed structure and the properties of these wave functions are described in 
Appendix B. 
A generic eigenstate of such system consists of $M$  ($1 \leq M \leq N$)  
"clusters" $\{\Omega_{\alpha}\}$ $(\alpha = 1,...,M)$ of bound particles. 
Each cluster is characterized by the momentum $q_{\alpha}$
of its center of mass motion, and by the number $n_{\alpha}$ of particles contained in it
(such that $\sum_{\alpha=1}^{M} n_{\alpha} = N$).  
Correspondingly, the eigenfunction $\Psi_{\bf q, n}^{(M)}(x_1,...,x_N)$ of such state
is characterized by $M$ continuous parameters ${\bf q} = (q_{1}, ..., q_{M})$  and $M$ 
integer parameters ${\bf n} = (n_{1}, ..., n_{M})$ (see Appendix B2, Eq.(\ref{C14})). The energy spectrum of this state is
\begin{equation}
\label{bas3-18}
E_{M}({\bf q,n})  = 
 \frac{1}{2\beta} \sum_{\alpha=1}^{M} \; n_{\alpha} q_{\alpha}^{2} 
- \frac{\kappa^{2}}{24\beta}\sum_{\alpha=1}^{M} (n_{\alpha}^{3}-n_{\alpha})
\end{equation}
where 
\begin{equation}
\label{bas3-18a}
\kappa \; = \; \beta^{3} u
\end{equation}
A general time dependent solution  $\Psi({\bf x},t)$
of the Schr\"odinger equation (\ref{bas2-7}) with the initial conditions, Eq.(\ref{bas2-8}),
can be represented in the form of the linear combination of the eigenfunctions
$\Psi_{\bf q, n}^{(M)}({\bf x})$ (Appendix B4, Eq.(\ref{C36})):
\begin{equation}
\label{bas3-19}
\Psi({\bf x},t) \; = \; \sum_{M=1}^{N} \;
{\sum_{{\bf n}}}' \; 
\int ' {\cal D} {\bf q} \; \; 
 \Psi_{\bf q, n}^{(M)}({\bf x}) \Psi_{\bf q, n}^{(M)^{*}}({\bf 0}) \; 
\exp\bigl[-E_{M}({\bf q,n}) \; t \bigr]
\end{equation}
where the summations over the integer parameters $n_{\alpha}$ and the integrations over the 
momenta $q_{\alpha}$ are performed in a restricted subspace, Eqs.(\ref{C28})-(\ref{C31}) and (\ref{C37}),  
which reflects the specific symmetry properties of the eigenfunctions 
$\Psi_{\bf q, n}^{(M)}({\bf x})$.
Correspondingly, according to Eq.(\ref{bas2-11}), for the replica partition function 
of the original directed polymer problem one gets (Appendix B4, Eq.(\ref{C45}))
\begin{equation}
\label{bas3-20}
Z(N,L) \; = \;  
\sum_{M=1}^{N} \; 
\frac{1}{M!}
\biggl[
\prod_{\alpha=1}^{M} 
\int_{-\infty}^{+\infty} \; \frac{dq_{\alpha}}{2\pi} 
\sum_{n_{\alpha}=1}^{\infty}
\biggr] \;
{\boldsymbol \delta}\biggl(\sum_{\alpha=1}^{M} n_{\alpha}, \; N \biggr) \;
\big| \Psi_{\bf q, n}^{(M)}({\bf 0}) \big|^2 \; 
\mbox{\LARGE e}^{-E_{M}({\bf q,n}) L}
\end{equation}
where due to the symmetry of the function 
$f\bigl({\bf q}, {\bf n}\bigr) = \bigl| \Psi_{\bf q, n}^{(M)}({\bf 0}) \bigr|^2 \; 
\exp\bigl[-E_{M}({\bf q,n}) \; L \bigr]$
with respect to permutations of all its $M$ pairs of arguments $(q_{\alpha}, n_{\alpha})$
the integrations over $M$ momenta $q_{\alpha}$ can be extended to the whole space $R_{M}$
while the summations over  $n_{\alpha}$'s are bounded by the only constrain
$\sum_{\alpha=1}^{M} n_{\alpha} = N$ (for simplicity, due to the presence
of the Kronecker symbol ${\boldsymbol \delta}\bigl(\sum_{\alpha} n_{\alpha}, \; N \bigr)$,
the summations over $n_{\alpha}$'s are extended to infinity).

\section{Replicas partition function and the free energy distribution function}

Using the explicit form of the wave functions $\Psi_{\bf q, n}^{(M)}({\bf x})$, Eq.(\ref{C14}),
the expression in Eq.(\ref{bas3-20}) for the replica partition function
can be reduced to (Appendix B4, Eq.(\ref{C47})-(\ref{C48}))
\begin{equation}
   \label{bas4-12}
 Z(N,L) \; =  \mbox{\LARGE e}^{-\beta N L f_{0}} \;\;
\tilde{Z}(N,\lambda) 
\end{equation}
where $f_{0} = \frac{1}{24}\beta^4 u^2 - \frac{1}{\beta L} \ln(\beta^{3} u)$ 
is the linear (selfaveraging) free energy density (cf. Eq.(\ref{bas1-10}), and
\begin{eqnarray}
\nonumber
\tilde{Z}(N.L) &=& 
N! \; \biggl\{
 \int_{-\infty}^{+\infty} \frac{dq}{2\pi\kappa N} \;
\exp\Bigl[
-\frac{L}{2\beta} N q^{2} + \frac{\kappa^{2}L}{24\beta} N^{3} 
\Bigr]
\; +
\\
\nonumber
\\
\nonumber
&+& 
\sum_{M=2}^{N} \frac{1}{M!} 
\biggl[
\prod_{\alpha=1}^{M}
\sum_{n_{\alpha}=1}^{\infty}
\int_{-\infty}^{+\infty} \frac{d q_{\alpha}}{2\pi\kappa n_{\alpha}} 
\biggr]
\;{\boldsymbol \delta}\biggl(\sum_{\alpha=1}^{M} n_{\alpha}, \; N \biggr) 
\prod_{\alpha<\beta}^{M} 
\frac{\big|q_{\alpha}-q_{\beta} -\frac{i\kappa}{2}(n_{\alpha}-n_{\beta})\big|^{2}}{
      \big|q_{\alpha}-q_{\beta} -\frac{i\kappa}{2}(n_{\alpha}+n_{\beta})\big|^{2}} 
\times
\\
&\times&
\exp\Bigl[
-\frac{L}{2\beta}\sum_{\alpha=1}^{M} n_{\alpha} q_{\alpha}^{2} + 
\frac{\kappa^{2}L}{24\beta} \sum_{\alpha=1}^{M} n_{\alpha}^{3}
\Bigr] 
\biggr\}
\label{bas4-13}
\end{eqnarray}
The first term in the above expression is the contribution of the ground state $(M=1)$,
while  the next terms $(M \geq 2)$ are the contributions of the rest of the energy
spectrum.

The terms cubic in
$n_{\alpha}$ in the exponential of Eq.\ (\ref{bas4-13})
can be linearised with the help of Airy
function, using the standard relation 
\begin{equation}
   \label{bas4-16}
\exp\bigl( \frac{1}{3} \lambda^{3} n^{3}\bigr) \; = \; 
\int_{-\infty}^{+\infty} dy \; \Ai(y) \; \exp(\lambda n)
\end{equation}
Redefining the momenta, $q_{\alpha} = \bigl(\beta\kappa/L\bigr)^{1/3} p_{\alpha}$ 
and introducing a new parameter
\begin{equation}
   \label{bas4-14}
\lambda (L) \; = \; \frac{1}{2} \biggl(\frac{L}{\beta} \kappa^{2} \biggr)^{1/3} \; = \; 
\frac{1}{2} \Bigl(\beta^{5} u^{2} L \Bigr)^{1/3}
\end{equation}
after shifting the Airy function parameters of integration
$y_{\alpha} \to y_{\alpha}  + p_{\alpha}^{2}$ the expression 
for $\tilde{Z}(N,\lambda)$ becomes sufficiently compact:
\begin{eqnarray}
\label{bas4-17}
\tilde{Z}(N,\lambda) &=& N! 
\int\int_{-\infty}^{+\infty} \frac{dy dp}{4\pi\lambda N} 
\; \Ai(y+p^{2}) \; \mbox{\LARGE e}^{\lambda N y}  + 
\\
\nonumber
 &+& N! \sum_{M=2}^{N} \frac{1}{M!} 
\biggl[\prod_{\alpha=1}^{M} \sum_{n_{\alpha}=1}^{\infty} \int\int_{-\infty}^{+\infty}
 \frac{dy_{\alpha} dp_{\alpha}}{4\pi\lambda n_{\alpha}} 
\Ai(y_{\alpha}+p^{2}_{\alpha}) \mbox{\LARGE e}^{\lambda n_{\alpha} y_{\alpha}}\biggr] 
\prod_{\alpha<\beta}^{M}
\frac{\big|\lambda(n_{\alpha}-n_{\beta}) -i(p_{\alpha}-p_{\beta})\big|^{2}}{
      \big|\lambda(n_{\alpha}+n_{\beta}) -i(p_{\alpha}-p_{\beta}) \big|^{2}} \;
{\boldsymbol \delta}\biggl(\sum_{\alpha=1}^{M}n_{\alpha}, \; N\biggr)
\end{eqnarray}
Now, using the Cauchy double alternant identity
\begin{equation}
 \label{tw4}
\frac{\prod_{\alpha<\beta}^{M} (a_{\alpha} - a_{\beta})(b_{\alpha} - b_{\beta})}{
     \prod_{\alpha,\beta=1}^{M} (a_{\alpha} - b_{\beta})} \; = \; 
(-1)^{M(M-1)/2} \det\Bigl[\frac{1}{a_{\alpha}-b_{\beta}}\Bigr]_{\alpha,\beta=1,...M}
\end{equation}
the product term in eq.(\ref{bas4-17}) can be represented in the determinant form:
\begin{equation}
\prod_{\alpha<\beta}^{M}
\frac{\big|\lambda(n_{\alpha}-n_{\beta}) -i(p_{\alpha}-p_{\beta})\big|^{2}}{
      \big|\lambda(n_{\alpha}+n_{\beta}) -i(p_{\alpha}-p_{\beta}) \big|^{2}}
 =  
\biggl[\prod_{\alpha=1}^{M} (2\lambda n_{\alpha}) \biggr]\; 
\det\Bigl[\frac{1}{\lambda n_{\alpha} -ip_{\alpha} 
                 + \lambda n_{\beta} + ip_{\beta}}\Bigr]_{\alpha,\beta=1,...M}
\label{tw5}
\end{equation}
Substituting now the expression for the replica partition function $\tilde{Z}(N,\lambda)$
into the definition of the probability function, eq.(\ref{tw3}), we can perform summation over $N$
(which would lift the constraint $\sum_{\alpha=1}^{M}n_{\alpha} =  N$) and obtain:
\begin{equation}
 \label{tw6}
W(x)  = 
\lim_{\lambda\to\infty} \Biggl\{
1  +  \sum_{M=1}^{\infty} \frac{(-1)^{M}}{M!}
\Biggl[\prod_{\alpha=1}^{M}
\int\int_{-\infty}^{+\infty}
\frac{dy_{\alpha} dp_{\alpha}}{2\pi}  
\Ai(y_{\alpha}+p^{2}_{\alpha})
\sum_{n_{\alpha}=1}^{\infty} (-1)^{n_{\alpha}-1} \mbox{\LARGE e}^{\lambda n_{\alpha} (y_{\alpha}+x)}
\Biggr] 
\det\Bigl[\frac{1}{\lambda n_{\alpha} -ip_{\alpha} + \lambda n_{\beta} + ip_{\beta}}\Bigr] \Biggr\}
\end{equation}
The above expression in nothing else but the expansion of the Fredholm determinant $\det(1 - \hat{K})$
(see e.g. \cite{Mehta})
with the kernel
\begin{equation}
\label{tw7}
\hat{K} \equiv
K\bigl[(n,p); (n',p')\bigr] = 
\Biggl[\int_{-\infty}^{+\infty} dy \Ai(y+p^{2}) (-1)^{n-1} \mbox{\LARGE e}^{\lambda n (y+x)}\Biggr]
\frac{1}{\lambda n -ip + \lambda n' + ip'} 
\end{equation}
Using the exponential representation of this determinant we get
\begin{equation}
 \label{tw8}
W(x)  = 
\lim_{\lambda\to\infty}
\exp\Bigl[-\sum_{M=1}^{\infty} \frac{1}{M} \; Tr \hat{K}^{M} \Bigr]
\end{equation}
where 
\begin{eqnarray}
\nonumber
Tr \hat{K}^{M} &=& 
\Biggl[\prod_{\alpha=1}^{M}
\int\int_{-\infty}^{+\infty}
\frac{dy_{\alpha} dp_{\alpha}}{2\pi}  
\Ai(y_{\alpha}+p^{2}_{\alpha})
\sum_{n_{\alpha}=1}^{\infty} (-1)^{n_{\alpha}-1} \mbox{\LARGE e}^{\lambda n_{\alpha} (y_{\alpha}+x)}
\Biggr] \times
\\
\nonumber
\\
&\times&
\frac{1}{(\lambda n_{1} -ip_{1} + \lambda n_{2} + ip_{2})
         (\lambda n_{2} -ip_{2} + \lambda n_{3} + ip_{3})...
(\lambda n_{M} -ip_{M} + \lambda n_{1} + ip_{1})}
 \label{tw9}
\end{eqnarray}
Substituting here
\begin{equation}
 \label{tw10}
\frac{1}{\lambda n_{\alpha} -ip_{\alpha} + \lambda n_{\alpha+1} + ip_{\alpha+1}}  =  
\int_{0}^{\infty} d\omega_{\alpha} 
\exp\bigl[-( \lambda n_{\alpha} - ip_{\alpha} + \lambda n_{\alpha+1} + ip_{\alpha+1}) \omega_{\alpha} \bigr]
\end{equation}
one can easily perform the summation over $n_{\alpha}$'s. Taking into account that
\begin{equation}
 \label{tw11}
\lim_{\lambda\to\infty} 
\sum_{n=1}^{\infty} (-1)^{n-1} \mbox{\LARGE e}^{\lambda n z} \; = \; 
\lim_{\lambda\to\infty} \frac{\mbox{\LARGE e}^{\lambda z}}{1 + \mbox{\LARGE e}^{\lambda z}}  =  
\theta(z)
\end{equation}
and shifting the integration parameters, $y_{\alpha} \to y_{\alpha} - x + \omega_{\alpha} + \omega_{\alpha-1}$
and $\omega_{\alpha} \to  \omega_{\alpha} + x/2$, after taking the thermodynamic limit,
$L \to \infty$ (which according to Eq.(\ref{bas4-14}) is equivalent to $\lambda \to \infty$) 
we obtain
\begin{equation}
\lim_{\lambda\to\infty} Tr \; \hat{K}^{M} = 
\prod_{\alpha=1}^{M}
\int_{0}^{\infty} dy_{\alpha} 
\int_{-\infty}^{+\infty} \frac{dp_{\alpha}}{2\pi}  
\int_{-x/2}^{\infty} d\omega_{\alpha} 
\Ai(y_{\alpha}+p^{2}_{\alpha}+\omega_{\alpha}+\omega_{\alpha-1}) \; 
\mbox{\LARGE e}^{ip_{\alpha} (\omega_{\alpha}-\omega_{\alpha-1})}
 \label{tw12}
\end{equation}
where by definition it is assumed that $\omega_{0} \equiv \omega_{M}$. Using the Airy function integral representation,
and taking into account that it satisfies the differential equation, $\Ai''(t) = t \Ai(t)$, 
one can easily perform the following integrations:
\begin{eqnarray}
\int_{0}^{\infty} dy 
\int_{-\infty}^{+\infty} \frac{dp}{2\pi}  
\Ai(y + p^{2} + \omega + \omega') 
\mbox{\LARGE e}^{ip (\omega-\omega')} &=&
 2^{-1/3} \int_{0}^{\infty} dy 
\Ai\bigl(2^{1/3} \omega + y\bigr) 
\Ai\bigl(2^{1/3} \omega' + y\bigr) 
\\
\nonumber
&=&
\frac{\Ai\bigl(2^{1/3} \omega\bigr) \Ai'\bigl(2^{1/3} \omega'\bigr) - 
      \Ai'\bigl(2^{1/3} \omega\bigr) \Ai\bigl(2^{1/3} \omega'\bigr)}{
\omega - \omega'}
\label{tw13}
\end{eqnarray}
Redefining $\omega_{\alpha} \to \omega_{\alpha} 2^{-1/3}$ we find
\begin{equation}
\lim_{\lambda\to\infty} Tr \hat{K}^{M} = 
\int\int ...\int_{-x/2^{2/3}}^{\infty} d\omega_{1} d\omega_{2} ... d\omega_{M}
K_{A}(\omega_{1},\omega_{2}) K_{A}(\omega_{2},\omega_{3}) ... K_{A}(\omega_{M},\omega_{1})
\label{tw14}
\end{equation}
where
\begin{equation}
 \label{tw15}
K_{A}(\omega,\omega') \; = \; 
\frac{\Ai(\omega) \Ai'(\omega') - \Ai'(\omega) \Ai(\omega')}{
\omega - \omega'}
\end{equation}
is the so called Airy kernel. This proves that in the thermodynamic limit, $L \to \infty$,
the probability function $W(x)$, eq.(\ref{tw1}), is defined by the Fredholm determinant,
\begin{equation}
 \label{tw16}
W(x)  \; = \; \det[1 - \hat{K}_{A}] \; \equiv \; F_{2}(-x/2^{2/3})
\end{equation}
where $\hat{K}_{A}$ is the integral operator on $[-x/2^{2/3}, \infty)$ with the 
Airy kernel, eq.(\ref{tw15}). The function $F_{2}(s)$ is the Tracy-Widom distribution \cite{Tracy-Widom}
\begin{equation}
 \label{tw17}
F_{2}(s) \; = \; \exp\Bigl(-\int_{s}^{\infty} dt \; (s-t) \; q^{2}(t)\Bigr)
\end{equation}
where the function $q(t)$ is the solution of the Panlev\'e II equation, $q'' = t q + 2 q^{3}$
with the boundary condition, $q(t\to +\infty) \sim Ai(t)$ (see Appendix C).

\section{Conclusions}

The Tracy-Widom distribution, Eq.(\ref{tw17}), was originally derived 
as the probability distribution of the
largest eigenvalue of a $n\times n$ random hermitian matrix in the limit $n \to \infty$.
At present there are exists an appreciable  list of statistical systems 
(which are not always look similar) in which the fluctuations of the quantities 
which play the role
of "energy"  are described by {\it the same} distribution function
$F_{2}(s)$. These systems are: 
the polynuclear growth (PNG) model \cite{PNG_Spohn},
the longest increasing subsequences (LIS) model \cite{LIS}, 
the longest common subsequences (LCS) \cite{LCS},
the oriented digital boiling model \cite{oriented_boiling}, 
the ballistic decomposition model \cite{ballistic_decomposition}, 
 the zero-temperature  lattice version of the directed polymers 
with an exponential and geometric site-disorder distribution \cite{DP_johansson}.
Now we can add to this list the continuous one-dimensional directed polymers with Gaussian
$\delta$-correlated random potential.

\newpage

\begin{center}

\appendix{\Large Appendix A}

\vspace{5mm}

 {\bf \large Quantum bosons with repulsive interactions}

\end{center}

\newcounter{A}
\setcounter{equation}{0}
\renewcommand{\theequation}{A.\arabic{equation}}

\vspace{5mm}

{\center{\bf 1. Eigenfunctions}}

\vspace{3mm}

The eigenstates equation 
for  $N$-particle system of one-dimensional quantum bosons with $\delta$-interactions is
\begin{equation}
 \label{A1}
\frac{1}{2}\sum_{a=1}^{N}\partial_{x_a}^2 \Psi({\bf x}) \; + \; 
\frac{1}{2}\kappa \sum_{a\not=b}^{N} \delta(x_a-x_b) \Psi({\bf x}) 
\; = \; - \beta E \Psi({\bf x})
\end{equation}
(where $\kappa = \beta^{3} u$). Due to the symmetry of the wave function 
with respect to permutations of its arguments
it is sufficient to consider it in the sector 
\begin{equation}
 \label{A2}
x_{1} \; < \; x_{2} \; < \; ... \; < \; x_{N}
\end{equation}
as well as at its boundary. Inside this sector the wave function $\Psi({\bf x})$ 
satisfy the equation
\begin{equation}
 \label{A3}
\frac{1}{2}\sum_{a=1}^{N}\partial_{x_a}^2 \Psi({\bf x}) 
\; = \; - \beta E \Psi({\bf x})
\end{equation}
which describes $N$ free particles, and its generic solution is the linear combination
of $N$ plane waves characterized by $N$ momenta $\{q_{1}, q_{2}, ..., q_{N}\} \equiv {\bf q}$.
Integrating Eq.(\ref{A1}) over the variable $(x_{i+1}-x_{i})$ in a small interval
around zero, $|x_{i+1}-x_{i}| < \epsilon \to  0$, and assuming that the other
variables $\{ x_{j}\}$ (with $j \not= i, i+1$) belong to the sector, Eq.(\ref{A2}),  
one easily finds that the wave function  $\Psi({\bf x})$ must satisfy the following
boundary conditions:
\begin{equation}
\label{A4}
\bigl(\partial_{x_{i+1}} - \partial_{x_i} + \kappa \bigr) 
\Psi({\bf x})\bigg|_{x_{i+1} = x_{i} + 0} \; = \; 0
\end{equation}
Functions satisfying both Eq.\ (\ref{A3}) and
the boundary conditions Eq.\ (\ref{A4}) can be written in the form
\begin{equation}
\label{A5}
\Psi_{q_{1}...q_{N}}(x_{1}, ..., x_{N}) \equiv
\Psi^{(N)}_{\bf q}({\bf x}) \; = \; C 
\biggl(\prod_{a<b}^{N}\bigl[ \partial_{x_a} - \partial_{x_b} + \kappa \bigr]\biggr) \;
    \det\Bigl[\exp( i q_c \, x_d) \Bigr]_{(c,d)=1,...,N}
\end{equation}
where $C$ is the normalization constant to be defined later.
First of all, it is evident that being the linear combination of the 
plane waves, the above wave function satisfy  Eq.(\ref{A3}). To demonstrate
which way this function satisfy the boundary conditions, Eq.(\ref{A4}),
let us check it, as an example, for the case $i=1$.
According to Eq.(\ref{A5}), the wave function $\Psi^{(N)}_{\bf q}({\bf x})$
can be represented in the form
\begin{equation}
\label{A6}
\Psi^{(N)}_{\bf q}({\bf x}) \; = \; 
- \bigl(\partial_{x_2} - \partial_{x_1} - \kappa \bigr) 
\tilde{\Psi}^{(N)}_{\bf q}({\bf x})
\end{equation}
where
\begin{equation}
\label{A7}
\tilde{\Psi}^{(N)}_{\bf q}({\bf x}) \; = \; C 
\biggl(\prod_{a=3}^{N}\bigl[ \partial_{x_1} - \partial_{x_a} + \kappa \bigr]
                      \bigl[ \partial_{x_2} - \partial_{x_a} + \kappa \bigr] \biggr)
\biggl(\prod_{3\leq a <b}^{N}\bigl[ \partial_{x_a} - \partial_{x_b} + \kappa \bigr]\biggr)
    \det\Bigl[\exp( i q_c \, x_d) \Bigr]_{(c,d)=1,...,N}
\end{equation}
One can easily see that this function is {\it antisymmetric} with respect to
the permutation of $x_{1}$ and $x_{2}$. Substituting Eq.(\ref{A6}) into Eq.(\ref{A4}) (with $i=1$)
we get
\begin{equation}
\label{A8}
-\biggl[\bigl(\partial_{x_2} - \partial_{x_1}\bigr)^{2} - \kappa^{2} \biggr] 
\tilde{\Psi}^{(N)}_{\bf q}({\bf x})\bigg|_{x_{2} = x_{1}} \; = \; 0
\end{equation}
Given the antisymmetry of the l.h.s expression with respect to
the permutation of $x_{1}$ and $x_{2}$ the above condition is indeed satisfied
at boundary $x_{1}=x_{2}$.

Since the eigenfunction $\Psi^{(N)}_{\bf q}({\bf x})$ satisfying 
Eq.(\ref{A1}) must be {\it symmetric} with respect to permutations of
its arguments, the function, Eq.(\ref{A5}), can be easily continued 
beyond the sector, Eq.(\ref{A2}), to the entire space of variables
$\{x_{1}, x_{2}, ..., x_{N}\} \in R_{N}$,
\begin{equation}
\label{A9}
\Psi^{(N)}_{\bf q}({\bf x}) \; = \; C 
\biggl(\prod_{a<b}^{N}\Bigl[ -i\bigl(\partial_{x_a} - \partial_{x_b}\bigr) +i \kappa \sgn(x_{a}-x_{b})\Bigr]\biggr) \;
    \det\Bigl[\exp( i  q_c \, x_d) \Bigr]_{(c,d)=1,...,N}
\end{equation}
where, by definition, the differential operators $\partial_{x_a}$ act only on 
the exponential terms and not on the $\sgn(x)$ functions, and 
for further convenience we have redefined $i^{N(N-1)/2} C \to C$. 
Explicitly the determinant in the above equation is
\begin{equation}
\label{A10}
     \det\Bigl[\exp( i  q_c \, x_d) \Bigr]_{(c,d)=1,...,N} \; = \; 
\sum_{P} (-1)^{[P]} \; \exp\Bigl[i \sum_{a=1}^{N} q_{p_{a}} x_{a} \Bigr]
\end{equation}
where the summation goes over the permutations $P$ of $N$ momenta $\{ q_{1}, q_{2}, ..., q_{N}\}$
over $N$ particles $\{ x_{1}, x_{2}, ..., x_{N}\}$, and $[P]$ denotes the parity of the permutation.
In this way the eigenfunction, Eq.(\ref{A9}), can be represented as follows
\begin{equation}
\label{A11}
\Psi^{(N)}_{\bf q}({\bf x}) \; = \; C 
\sum_{P} (-1)^{[P]} \; 
\biggl(\prod_{a<b}^{N}\Bigl[ -i\bigl(\partial_{x_a} - \partial_{x_b}\bigr) +i \kappa \sgn(x_{a}-x_{b})\Bigr]\biggr) \;
\exp\Bigl[i \sum_{a=1}^{N} q_{p_{a}} x_{a} \Bigr]
\end{equation}
 Taking the derivatives, we obtain
\begin{equation}
\label{A12}
\Psi^{(N)}_{\bf q}({\bf x}) \; = \; C 
\sum_{P} (-1)^{[P]} \; 
\biggl(\prod_{a<b}^{N}\Bigl[ q_{p_a} - q_{p_b} +i \kappa \sgn(x_{a}-x_{b})\Bigr]\biggr) \;
\exp\Bigl[i \sum_{a=1}^{N} q_{p_{a}} x_{a} \Bigr]
\end{equation}
It is evident from these representations that the eigenfunctions $\Psi^{(N)}_{\bf q}({\bf x})$ are
{\it antisymmetric} with respect to permutations of the momenta $q_{1}, ..., q_{N}$. 

Finally, substituting the expression for the eigenfunctions, Eq.(\ref{A5}) (which is valid in
the sector, Eq.(\ref{A2})), into Eq.(\ref{A3}) for the energy spectrum we find
\begin{equation}
   \label{A20}
E \;  = \; \frac{1}{2\beta} \sum_{a=1}^{N} q_{a}^{2} 
\end{equation}

\vspace{10mm}

{\center{\bf 2. Orthonormality}}

\vspace{3mm}

Now one can easily prove that the above eigenfunctions with different momenta are orthogonal to
each other. Let us consider two wave functions $\Psi^{(N)}_{\bf q}({\bf x})$ and 
$\Psi^{(N)}_{\bf q'}({\bf x})$ where it is assumed that
\begin{eqnarray}
 \label{A13}
q_{1} \; < \; q_{2} \; < \; ... \; < \; q_{N}\\
\nonumber
q'_{1} \; < \; q'_{2} \; < \; ... \; < \; q'_{N}
\end{eqnarray}
Using the representation, Eq.(\ref{A11}), for the overlap of these two function we get
\begin{eqnarray}
 \label{A14}
\nonumber
\overline{\Psi^{(N)^{*}}_{\bf q'}({\bf x}) \Psi^{(N)}_{\bf q}({\bf x})} &\equiv& 
\int_{-\infty}^{+\infty} d^{N}{\bf x} \; \Psi^{(N)^{*}}_{\bf q'}({\bf x}) \Psi^{(N)}_{\bf q}({\bf x}) \\
\nonumber
&=& |C|^{2} \sum_{P,P'} (-1)^{[P]+[P']}  
\int_{-\infty}^{+\infty} d^{N}{\bf x} 
\biggl\{\biggl(\prod_{a<b}^{N}
\bigl[ i\bigl(\partial_{x_a} - \partial_{x_b}\bigr) -i \kappa \sgn(x_{a}-x_{b})\bigr]\biggr) \;
\exp\bigl[-i \sum_{a=1}^{N} q'_{p'_{a}} x_{a} \bigr]\biggr\} \times \\
&\times& 
\biggl\{\biggl(\prod_{a<b}^{N}
\bigl[ -i\bigl(\partial_{x_a} - \partial_{x_b}\bigr) +i \kappa \sgn(x_{a}-x_{b})\bigr]\biggr) \;
\exp\bigl[i \sum_{a=1}^{N} q_{p_{a}} x_{a} \bigr]\biggr\}
\end{eqnarray}
Integrating by parts we obtain
\begin{eqnarray}
 \label{A15}
\overline{\Psi^{(N)^{*}}_{\bf q'}({\bf x}) \Psi^{(N)}_{\bf q}({\bf x})} &=& |C|^{2}
\sum_{P,P'} (-1)^{[P]+[P']}  
\int_{-\infty}^{+\infty} d^{N}{\bf x} \;
\exp\bigl[-i \sum_{a=1}^{N} q'_{p'_{a}} x_{a} \bigr] \times \\
\nonumber
&\times&
\biggl(\prod_{a<b}^{N}\bigl[-i\bigl(\partial_{x_a} - \partial_{x_b}\bigr) -i \kappa \sgn(x_{a}-x_{b})\bigr] \;
                      \bigl[-i\bigl(\partial_{x_a} - \partial_{x_b}\bigr) +i \kappa \sgn(x_{a}-x_{b})\bigr] \biggr)
\exp\bigl[i \sum_{a=1}^{N} q_{p_{a}} x_{a} \bigr]
\end{eqnarray}
or
\begin{equation}
 \label{A16}
\overline{\Psi^{(N)^{*}}_{\bf q'}({\bf x}) \Psi^{(N)}_{\bf q}({\bf x})} = |C|^{2}
\sum_{P,P'} (-1)^{[P]+[P']}  
\int_{-\infty}^{+\infty} d^{N}{\bf x} 
\exp\bigl[-i \sum_{a=1}^{N} q'_{p'_{a}} x_{a} \bigr]
\biggl(\prod_{a<b}^{N}\bigl[-(\partial_{x_a} - \partial_{x_b})^{2} + \kappa^{2} \bigr] \biggr)
\exp\bigl[i \sum_{a=1}^{N} q_{p_{a}} x_{a} \bigr]
\end{equation}
Taking the derivatives and performing the integrations
we find
\begin{eqnarray}
 \label{A17}
\nonumber
\overline{\Psi^{(N)^{*}}_{\bf q'}({\bf x}) \Psi^{(N)}_{\bf q}({\bf x})} &=& 
|C|^{2} \sum_{P,P'} (-1)^{[P]+[P']}  
\biggl(\prod_{a<b}^{N}\bigl[(q_{p_a} - q_{p_b})^{2} + \kappa^{2} \bigr] \biggr)
\int_{-\infty}^{+\infty} d^{N}{\bf x} \; 
\exp\bigl[i \sum_{a=1}^{N} (q_{p_{a}}-q'_{p'_{a}}) x_{a} \bigr] \\ 
&=& 
|C|^{2} \sum_{P,P'} (-1)^{[P]+[P']} 
\biggl(\prod_{a<b}^{N}\bigl[(q_{p_a} - q_{p_b})^{2} + \kappa^{2} \bigr] \biggr)
\biggl[\prod_{a=1}^{N} (2\pi) \delta(q_{p_a} - q'_{p'_a}) \biggr]
\end{eqnarray}
Taking into account the constraint, Eq.(\ref{A13}), one can easily note that the only the terms
which survive in the above summation over the permutations are $P = P'$, all contributing
equal value. Thus, we finally get
\begin{equation}
 \label{A18}
\overline{\Psi^{(N)^{*}}_{\bf q'}({\bf x}) \Psi^{(N)}_{\bf q}({\bf x})} = |C|^{2} \; N! \;
\biggl(\prod_{a<b}^{N}\bigl[(q_{a} - q_{b})^{2} + \kappa^{2} \bigr] \biggr)
\biggl[\prod_{a=1}^{N} (2\pi) \delta(q_{a} - q'_{a}) \biggr]
\end{equation}
With the normalization constant
\begin{equation}
   \label{A19}
\big|C({\bf q})\big|^{2} = \frac{1}{N! \prod_{a<b}^{N} 
\bigl[ (q_{a} - q_{b})^{2} + \kappa^{2} \bigr] }
\end{equation}
we conclude that the set of the eigenfunctions, Eq.(\ref{A11}) or (\ref{A12}), are
orthonormal. The proof of completeness of this set is given in Ref. \cite{gaudin}.
It should be noted that the above wave functions present the orthonormal set of
eigenfunctions of the problem, Eq.(\ref{A1}), for any sign of the interactions
$\kappa$, e.i. both for the repulsive, $\kappa < 0$, and for the attractive, $\kappa > 0$, cases.
However, only in the case of repulsion this set is complete, while in the 
case of attractive interactions, $\kappa > 0$, in addition to the solutions, Eq.(\ref{A11}),
which describe the continuous free particles spectrum, one finds the whole family
of discrete bound eigenstates (which do not exist in the case of repulsion).


\vspace{10mm}

\begin{center}

\appendix{\Large Appendix B}

\vspace{5mm}

 {\bf \large Quantum bosons with attractive interactions}

\end{center}

\newcounter{B}
\setcounter{equation}{0}
\renewcommand{\theequation}{B.\arabic{equation}}

\vspace{5mm}

{\center{\bf 1. Ground state}}

\vspace{3mm}

The simplest example of the bound eigenstate defined by eq.(\ref{A1}) (with $\kappa > 0$)
is the one in which all $N$ particles are bound into a single "cluster":
\begin{equation}
   \label{B1}
   \Psi_{q}^{(1)}({\bf x}) \; = \;  C \; 
    \exp\biggl[i q \sum_{a=1}^{N} x_{a} - \frac{1}{4}\kappa \sum_{a,b=1}^{N} |x_{a}-x_{b}| \biggr]
\end{equation}
where $C$ is the normalization constant (to be defined below) and
$q$ is the continuous momentum of free center of mass motion. Substituting
this function in Eq.(\ref{A1}), one can easily check that  this is indeed the eigenfunction
with the energy spectrum given by the relation
\begin{equation}
 \label{B2}
E \;  = \;  -\frac{1}{2\beta} \sum_{a=1}^{N} 
\biggl[iq - \frac{1}{2}\kappa \sum_{b=1}^{N} \sgn(x_{a}-x_{b}) \biggr]^{2}
\end{equation}
where it is assumed (by definition) that $\sgn(0) = 0$.
Since the result of the above summations does not depend on the mutual particles positions,
for simplicity we can order them according to Eq.(\ref{A2}). Then, using well known relations
\begin{eqnarray}
 \label{B3}
\sum_{b=1}^{N} \sgn(x_{a}-x_{b}) &=& -(N+1-2a)\\
\label{B4}
\sum_{a=1}^{N} a &=& \frac{1}{2} N (N+1)\\
\label{B5}
\sum_{a=1}^{N} a^{2} &=& \frac{1}{6} (N+1) (2N+1)
\end{eqnarray} 
for the energy spectrum, Eq.(\ref{B2}), we get
\begin{equation}
   \label{B6}
E \; = \; \frac{N}{2\beta} q^2 - \frac{\kappa^{2}}{24\beta}(N^{3}-N) \; \equiv \; 
E_{1}(q, N)
\end{equation}
The normalization constant $C$ is defined by the orthonormality
condition
\begin{equation}
   \label{B7}
\overline{\Psi_{q'}^{(1)^{*}}({\bf x}) \Psi_{q}^{(1)}({\bf x})} \; \equiv \;
\int_{-\infty}^{+\infty} dx_{1}...dx_{N} \; 
\Psi_{q'}^{(1)^{*}}({\bf x}) \Psi_{q}^{(1)}({\bf x}) \; = \; 
(2\pi) \delta(q-q')
\end{equation}
Substituting here Eq.(\ref{B1}) we get
\begin{eqnarray}
 \nonumber
\overline{\Psi_{q'}^{(1)^{*}}({\bf x}) \Psi_{q}^{(1)}({\bf x})} &=&
|C|^{2} \int_{-\infty}^{+\infty} dx_{1}...dx_{N} \; 
\exp\biggl[i (q-q') \sum_{a=1}^{N} x_{a} - 
\frac{1}{2}\kappa \sum_{a,b=1}^{N} |x_{a}-x_{b}| \biggr] 
\\
\label{B8}
&=& |C|^{2} N!
\int_{-\infty}^{+\infty} dx_{1}
\int_{x_{1}}^{+\infty} dx_{2} 
....
\int_{x_{N-1}}^{+\infty} dx_{N}
\exp\biggl[i (q-q') \sum_{a=1}^{N} x_{a} + \kappa \sum_{a=1}^{N} (N+1-2a) x_{a} \biggr]
\end{eqnarray}
where for the ordering, Eq.(\ref{A2}), we have used the relation
\begin{equation}
 \label{B9}
\frac{1}{2} \sum_{a,b=1} |x_{a}-x_{b}| \; = \; 
- \sum_{a=1}^{N} (N+1-2a) x_{a}
\end{equation}
Integrating first over $x_{N}$, then over $x_{N-1}$, and proceeding until $x_{1}$,
we find
\begin{eqnarray}
 \nonumber
\overline{\Psi_{q'}^{(1)^{*}}({\bf x}) \Psi_{q}^{(1)}({\bf x})} 
&=& |C|^{2} N! \;
\biggl(\prod_{r=1}^{N-1} \frac{1}{r[(N-r)\kappa -i(q-q')]}\biggr)
\int_{-\infty}^{+\infty} dx_{1} 
\exp\bigl[iN(q-q')x_{1}\bigr]\\
\nonumber
&=& |C|^{2} N! \; 
\biggl(\prod_{r=1}^{N-1} \frac{1}{r(N-r)\kappa}\biggr) \;
(2\pi) \delta\bigl(N(q-q')\bigr) \\
\label{B10}
&=& |C|^{2} \; \frac{N\kappa}{N!\kappa^{N}} \; (2\pi) \delta(q-q')
\end{eqnarray}
According to Eq.(\ref{B7}) this defines the normalization constant
\begin{equation}
   \label{B11}
   C \; = \; \sqrt{\frac{\kappa^N N!}{\kappa N}}  \; \equiv \;   C^{(1)}(q)
\end{equation}
Note that the eigenstate described by the considered wave function, Eq.(\ref{B1}), exists only
in the case of attraction, $\kappa > 0$, otherwise this function is divergent at infinity 
and consequently it is not normalizable.

\vspace{5mm}

It should be noted that the wave function, Eq.(\ref{B1}), can also be derived from the 
general eigenfunctions structure, Eq.(\ref{A12}), by introducing (discrete) imaginary 
parts for the momenta $q_{a}$. We assume again that the position of particles 
are ordered according to Eq.(\ref{A2}), and  define the particles' momenta
according to the rule
\begin{equation}
\label{B12}
q_{a} \; = \; q - \frac{i}{2} \kappa (N+1-2a)
\end{equation}
Substituting this into Eq.(\ref{A12}) we get
\begin{eqnarray}
  \nonumber 
  \Psi_{q}^{(1)}({x_{1} < x_{2} < ... < x_{N}}) & \propto & 
\sum_{P} (-1)^{[P]} \; 
\biggl(
\prod_{a<b}^{N}
\biggl[ 
  \Bigl( q - \frac{i}{2} \kappa (N+1-2P_{a})\Bigr) 
- \Bigl( q - \frac{i}{2} \kappa (N+1-2P_{b})\Bigr) -i \kappa 
\biggr]
\biggr) \times
\\
&&\times
\exp\biggl[ iq\sum_{a=1}^{N}x_{a} +\frac{\kappa}{2}\sum_{a=1}^{N} (N+1-2P_{a}) x_{a} \biggr]  
\\
\nonumber
\\
& \propto &
\sum_{P} (-1)^{[P]} 
\biggl(\prod_{a<b}^{N}\bigl[ P_{b}-P_{a} + 1 \bigr] \biggr) \;
\exp\biggl[ iq\sum_{a=1}^{N}x_{a} +\frac{\kappa}{2}\sum_{a=1}^{N} (N+1-2P_{a}) x_{a} \biggr] 
\label{B13}
\end{eqnarray}
Here one can easily note that due to the presence of the product 
$\prod_{a<b}^{N}[P_{b}-P_{a} + 1] $ in the summation over permutations only the 
trivial one, $P_{a} = a$, gives non-zero contribution (if we permute
any two numbers in the sequence $1, 2, ... , N$ then we can always find two numbers
$a < b$, such that $P_{b} = P_{a} - 1$). Thus
\begin{equation}
 \label{B14}
 \Psi_{q}^{(1)}({x_{1} < x_{2} < ... < x_{N}}) \; \propto \;
\exp\biggl[ iq\sum_{a=1}^{N}x_{a} +\frac{\kappa}{2}\sum_{a=1}^{N} (N+1-2a) x_{a} \biggr] 
\end{equation}
Taking into account the relation, Eq.(\ref{B9}), we recover the function, Eq.(\ref{B1}),
which is symmetric with respect to its $N$ arguments and therefore can be extended
beyond the sector, Eq.(\ref{A2}), for arbitrary particles positions.
Finally, substituting the momenta, Eq.(\ref{B12}), into the general expression for the energy 
spectrum, Eq.(\ref{A20}), we get
\begin{equation}
 \label{B15}
E \, = \; \frac{1}{2\beta} \sum_{a=1}^{N} \bigl[q - \frac{i}{2} \kappa (N+1-2a) \bigr]^{2} 
\end{equation}
Performing here simple summations (using Eqs.(\ref{B4}), (\ref{B5})) 
one recovers Eq.(\ref{B6}).

\vspace{10mm}

{\center{\bf 2. Eigenfunctions}}

\vspace{5mm}

A generic eigenfunction of attractive bosons is characterized by $N$ momenta parameters 
$\{ q_{a}\} \; (a= 1,2,...N)$ which  may have imaginary parts.
It is convenient to group these parameters into $M \; \; (1 \leq M \leq N)$ "vector" momenta, 
\begin{equation}
\label{C1}
q^{\alpha}_{r} \; = \; 
q_{\alpha} \; - \; \frac{i}{2} \; \kappa \; (n_{\alpha} + 1 -2 r)
\end{equation}
where $q_{\alpha} \; (\alpha = 1, 2, ..., M)$ are the continuous (real) parameters,
and the discrete imaginary components of each "vector"  are labeled by an index 
$r = 1, 2, ..., n_{\alpha}$. With the given total number of particles equal to
$N$, the integers $n_\alpha$ have to satisfy the constraint
\begin{equation}
\label{C2}
\sum_{\alpha = 1}^{M} \; n_{\alpha} \; = \; N
\end{equation}
In other words, a generic eigenstate is characterized by the discrete number $M$
of complex ''vector`` momenta, by the set
of $M$ integer parameters $\{ n_{1}, n_{2}, ..., n_{M} \} \equiv {\bf n}$ (which are the numbers
of imaginary components of each ''vector``) and by the set of $M$ real continuous
momenta $\{ q_{1}, q_{2}, ..., q_{M} \}  \equiv {\bf q}$. 

The general expression for the eigenfunctions is given in Eqs.(\ref{A9})-(\ref{A12}).
To understand the structure of the determinant of the $N\times N$ matrix
$\exp( i  q_{a}  x_{b})$, which defines these wave functions, 
the $N$ momenta $q_{a}$, eq.(\ref{C1}), can be ordered as follows:
\begin{equation}
 \label{C3}
\{ q_{a} \} \; \equiv \; \{q^{\alpha}_{r}\} \; = \; 
                    \{ q^{1}_{1}, \; q^{1}_{2}, \; ... \; , \; q^{1}_{n_{1}}; \; 
                       q^{2}_{1}, \; q^{2}_{2}, \; ... \; , \; q^{2}_{n_{2}}; \; ... \; ; \; 
                       q^{M}_{1}, \; q^{M}_{2}, \; ... \; , \; q^{M}_{n_{M}} \}
\end{equation}
By definition,
\begin{equation}
\label{C4}
    \det\Bigl[\exp( i q_a \, x_c)\Bigr]_{(c,d)=1,...,N} \; = \; 
\sum_{P} (-1)^{[P]} \; \exp\Bigl[i \sum_{a=1}^{N} q_{p_{a}} x_{a} \Bigr]
\end{equation}
where the summation goes over the permutations of $N$ momenta $\{ q_{a}\}$, Eq.(\ref{C3}),
over $N$ particles $\{ x_{1}, x_{2}, ..., x_{N}\}$, and $[P]$ denotes the parity of the permutation.
For a given permutation $P$ a particle number $a$ 
is attributed a momentum component $q^{\alpha(a)}_{r(a)}$.
The particles getting the momenta with the same 
 $\alpha$ (having the same real part $q_{\alpha}$) will be called  
belonging to a cluster $\Omega_{\alpha}$. For a given permutation $P$
the particles belonging 
to the same cluster are  numbered by the  "internal" index $r = 1,...,n_{\alpha}$.
Thus, according to Eq.(\ref{A11}),
\begin{equation}
 \label{C5}
\Psi_{\bf q, n}^{(M)}({\bf x}) \; = \;  C^{(M)}_{\bf q, n}
\sum_{P} (-1)^{[P]} 
    \biggl(\prod_{a<b}^{N}\Bigl[ -i\bigl(\partial_{x_a} - \partial_{x_b}\bigr) +i \kappa \sgn(x_{a}-x_{b})\Bigr]\biggr) \;
    \exp\biggl[ i \sum_{c=1}^{N} q^{\alpha(c)}_{r(c)} \, x_{c} \biggr]
\end{equation}
where $C^{(M)}_{\bf q, n}$ is the normalization constant to be defined later.
Substituting here Eq.(\ref{C1}) and taking derivatives we get
\begin{eqnarray}
   \label{C6}
\nonumber
  \Psi_{\bf q, n}^{(M)}({\bf x}) &=& C^{(M)}_{\bf q, n}
\sum_{P} (-1)^{[P]} 
\prod_{a<b}^{N}
\biggl[ 
  \Bigl(q_{\alpha(a)} - \frac{i\kappa}{2} \big[n_{\alpha(a)} + 1 - 2r(a)\big] \Bigr)
- \Bigl(q_{\alpha(b)} - \frac{i\kappa}{2} \big[n_{\alpha(b)} + 1 - 2r(b)\big] \Bigr) 
+i\kappa \sgn(x_{a}-x_{b})
\biggr]  
\times\\
& &\times\exp\biggl[
i\sum_{c=1}^{N}q_{\alpha(c)} x_{c} 
 +\frac{\kappa}{2}\sum_{c=1}^{N} \Bigl(n_{\alpha(c)} +1 - 2r(c)\Bigr) x_{c} \biggr]
\end{eqnarray}
The pre-exponential product in the above equation contains two types of term: the pairs of 
points $(a,b)$ which belong to different clusters ($\alpha(a)\not=\alpha(b)$), and pairs of points
which belong to the same cluster ($\alpha(a) = \alpha(b)$). In the last case, the product
$\Pi_{\alpha}$ over the pairs of points which belong to a cluster $\Omega_{\alpha}$ 
reduces to 
\begin{equation}
 \label{C7}
\Pi_{\alpha} \; \propto \; 
\prod_{a<b\in\Omega_{\alpha}}\bigl[r(b) - r(a) -\sgn(x_{a}-x_{b})]\bigr] 
\end{equation}
As for the ground state wave function Eq.\ (\ref{B13})--(\ref{B14}),
one can easily note that due to the presence of this product 
in the summations over $n_{\alpha}!$ ''internal`` (inside the cluster $\Omega_{\alpha}$)
permutations $r(a)$ 
only one permutation gives non-zero contribution. 
To prove this statement, we note
that the wave function $\Psi_{\bf q, n}^{(M)}({\bf x})$ is symmetric with
respect to permutations of its $N$ arguments $\{x_{a}\}$; it is then
sufficient to consider the case where the positions of the particles are
ordered, $x_1 < x_2 < \dots < x_N$. In particular,
the particles $\{x_{a_{k}}\} \; (k = 1, 2, \dots, n_{\alpha})$
belonging to the same cluster $\Omega_{\alpha}$ are also ordered $x_{a_{1}} <
x_{a_{2}} < \dots < x_{a_{n_{\alpha}}}$.  In this case
\begin{equation}
 \label{C8}
\Pi_{\alpha} \; \propto \; 
\prod_{k<l}^{n_{\alpha}} \bigl[r(l) - r(k) + 1\bigr] 
\end{equation}
Now it is evident that the above product is non-zero only for the trivial
permutation, $r(k) = k$ (since if we permute
any two numbers in the sequence $1, 2, ... , n_{\alpha}$, we can always find two numbers
$k < l$, such that $r(l) = r(k) - 1$). In this case
\begin{equation}
 \label{C9}
\Pi_{\alpha} \; \propto \; 
\prod_{k<l}^{n_{\alpha}} \bigl[ l - k + 1 \bigr] 
\end{equation}
Including the values of all these "internal" products, 
Eq.(\ref{C9}), into the redefined normalization constant
$C^{(M)}_{\bf q, n}$, for the wave function, Eq.(\ref{C6}) 
(with  $x_1 < x_2 < \dots < x_N$), we obtain
\begin{eqnarray}
   \label{C10}
\nonumber
  \Psi_{\bf q, n}^{(M)}({\bf x}) &=& C^{(M)}_{\bf q, n}
{\sum_{P}}' (-1)^{[P]} 
\prod_{\substack{ a<b\\ \alpha(a)\not=\alpha(b) }}^{N}  
\biggl[ 
\Bigl(q_{\alpha(a)} - \frac{i\kappa}{2} n_{\alpha(a)}\Bigr) -
\Bigl(q_{\alpha(b)} - \frac{i\kappa}{2} n_{\alpha(b)}\Bigr) 
+i\kappa\Bigl(r(a) - r(b) - 1)\Bigr)\Biggr]  
\times\\
& &\times\exp\biggl[
i\sum_{c=1}^{N}q_{\alpha(c)} x_{c} 
 +\frac{\kappa}{2}\sum_{c=1}^{N} \Bigl(n_{\alpha(c)} +1 - 2r(c)\Bigr) x_{c} \biggr]
\end{eqnarray}
where  the product  goes now only 
 over the pairs of particles belonging to {\it different}
clusters, and the symbol ${\sum_{P}}'$ means 
that the summation goes only over 
the permutations $P$ in which the "internal" indices
$r(a)$ are ordered inside each cluster.
Note that although the positions of particles belonging to the same cluster are 
ordered, the mutual positions of particles belonging to different clusters could be arbitrary,
so that geometrically the clusters are free to "penetrate" each other. In other words, 
the name "cluster" does to assume geometrically compact particles positions. 

Now taking into account the symmetry of the wave function $\Psi_{\bf q, n}^{(M)}({\bf x})$
with respect to the permutations of its arguments the expression in Eq.(\ref{C10}) 
can be easily continued beyond the
the sector $x_1 < x_2 < ... < x_N$ for the entire coordinate space $R_{N}$.
Using the relations
\begin{equation}
 \label{C11}
\sum_{a\in\Omega_{\alpha}} \bigl(n_{\alpha} +1 - 2r(a)\bigr) x_{a} \; = \; 
\sum_{k=1}^{n_{\alpha}} (n_{\alpha} +1 - 2 k) x_{a_{k}} \; = \; 
-\frac{1}{2} \sum_{k,l=1}^{n_{\alpha}}
\big|x_{a_{k}} - x_{a_{l}}\big|
\end{equation}
(where $x_{a_{1}} < x_{a_{2}} < ... < x_{a_{n_{\alpha}}}$), 
for the wave function $\Psi_{\bf q, n}^{(M)}({\bf x})$ with {\it arbitrary} particles positions 
we get the following sufficiently compact representation (cf. Eq.(\ref{C5})):
\begin{equation}
 \label{C14}
 \Psi_{\bf q, n}^{(M)}({\bf x}) =  C^{(M)}_{\bf q, n}
{\sum_{P}}' (-1)^{[P]} 
\prod_{\substack{ a<b\\ \alpha(a)\not=\alpha(b) }}^{N} 
\biggl[ -i\bigl(\partial_{x_a} - \partial_{x_b}\bigr) 
+ i \kappa \sgn(x_{a}-x_{b})\biggr]
 \exp\biggl[
i\sum_{\alpha=1}^{M}q_{\alpha}\sum_{c\in\Omega_{\alpha}}^{n_{\alpha}} x_{c} 
 -\frac{\kappa}{4}\sum_{\alpha=1}^{M}\sum_{c,c'\in\Omega_{\alpha}}^{n_{\alpha}} |x_{c}-x_{c'}| \biggr]
\end{equation}
Finally, substituting Eq.(\ref{C1})-(\ref{C2}) into Eq.(\ref{A20}), 
for the energy spectrum one easily obtains:
\begin{equation}
\label{C14a}
E_{M}({\bf q,n}) \; = \;
\frac{1}{2\beta} \sum_{\alpha=1}^{M} \; \sum_{r=1}^{n_{\alpha}} (q^{\alpha}_{k})^{2} 
\; = \; \frac{1}{2\beta} \sum_{\alpha=1}^{M} \; n_{\alpha} q_{\alpha}^{2} \;
- \; \frac{\kappa^{2}}{24\beta}\sum_{\alpha=1}^{M} (n_{\alpha}^{3}-n_{\alpha})
\end{equation}

\vspace{10mm}

{\center {\bf 3. Orthonormality}}

\vspace{3mm}

We define the overlap of two wave functions characterized by two sets of
parameters, $(M, {\bf n}, {\bf q})$ and $(M', {\bf n'}, {\bf q'})$ as
\begin{equation}
\label{C15}
Q^{(M,M')}_{{\bf n},{\bf n'}}({\bf q}, {\bf q'}) \; \equiv \; 
\int_{-\infty}^{+\infty} d^{N}{\bf x} \; 
 \Psi_{\bf q', n'}^{(M')^{*}}({\bf x})
 \Psi_{\bf q, n}^{(M)}({\bf x})  
\end{equation}
Substituting here Eq.(\ref{C14}) we get
\begin{eqnarray}
\label{C16}
Q^{(M,M')}_{{\bf n},{\bf n'}}({\bf q}, {\bf q'}) &=&
C^{(M)}_{\bf q, n} C^{(M')^{*}}_{\bf q', n'} 
{\sum_{P}}' {\sum_{P'}}' (-1)^{[P] + [P']}  \; 
\int_{-\infty}^{+\infty} d^{N}{\bf x}  
\\
\nonumber
\\
\nonumber
&& \times
\biggl(
\prod_{\substack{ a<b\\ \alpha'(a)\not=\alpha'(b) }}^{N} \biggl[ 
i \bigl(\partial_{x_a} - \partial_{x_b}\bigr) - i \kappa \sgn(x_{a}-x_{b}) \biggr] \biggr)
\exp\biggl[
-i\sum_{\alpha=1}^{M'}q'_{\alpha}\sum_{c\in\Omega'_{\alpha}}^{n'_{\alpha}} x_{c} 
 -\frac{\kappa}{4}\sum_{\alpha=1}^{M'}\sum_{c,c'\in\Omega'_{\alpha}}^{n'_{\alpha}} |x_{c}-x_{c'}| \biggr]
\times
\\
\nonumber
\\
\nonumber
&& \times
\biggl(
\prod_{\substack{ a<b\\ \alpha(a)\not=\alpha(b) }}^{N} \biggl[ 
-i \bigl(\partial_{x_a} - \partial_{x_b}\bigr) +i \kappa \sgn(x_{a}-x_{b}) \biggr] \biggr)
\exp\biggl[
i\sum_{\alpha=1}^{M}q_{\alpha}\sum_{c\in\Omega_{\alpha}}^{n_{\alpha}} x_{c} 
 -\frac{\kappa}{4}\sum_{\alpha=1}^{M}\sum_{c,c'\in\Omega_{\alpha}}^{n_{\alpha}} |x_{c}-x_{c'}| \biggr]
\end{eqnarray}
where $\{\Omega_{\alpha}\}$ and $\{\Omega'_{\alpha}\}$ denote the clusters of the permutations $P$
and $P'$ correspondingly. Integrating by parts we obtain
\begin{eqnarray}
\nonumber
Q^{(M,M')}_{{\bf n},{\bf n'}}({\bf q}, {\bf q'}) &=&
C^{(M)}_{\bf q, n} C^{(M')^{*}}_{\bf q', n'} 
{\sum_{P}}' {\sum_{P'}}' (-1)^{[P] + [P']}  
\int_{-\infty}^{+\infty} d^{N}{\bf x}  \; 
\exp\biggl[
-i\sum_{\alpha=1}^{M'}q'_{\alpha}\sum_{c\in\Omega'_{\alpha}}^{n'_{\alpha}} x_{c} 
 -\frac{\kappa}{4}\sum_{\alpha=1}^{M'}\sum_{c,c'\in\Omega'_{\alpha}}^{n'_{\alpha}} |x_{c}-x_{c'}| \biggr] \times
\\
\nonumber
\\
\nonumber
&& \times
\biggl(
\prod_{\substack{ a<b\\ \alpha'(a)\not=\alpha'(b) }}^{N} \biggl[ 
-i\bigl(\partial_{x_a} - \partial_{x_b}\bigr) - i \kappa \sgn(x_{a}-x_{b}) \biggr] \biggr)
\biggl(
\prod_{\substack{ a<b\\ \alpha(a)\not=\alpha(b) }}^{N} \biggl[ 
-i\bigl(\partial_{x_a} - \partial_{x_b}\bigr) +i \kappa \sgn(x_{a}-x_{b}) \biggr] \biggr) \times
\\
\nonumber
\\
&& \times
\exp\biggl[
i\sum_{\alpha=1}^{M}q_{\alpha}\sum_{c\in\Omega_{\alpha}}^{n_{\alpha}} x_{c} 
 -\frac{\kappa}{4}\sum_{\alpha=1}^{M}\sum_{c,c'\in\Omega_{\alpha}}^{n_{\alpha}} |x_{c}-x_{c'}| \biggr]
\label{C17}
\end{eqnarray}

First, let us consider the case when the integer parameters of the two functions
coincide, $M = M'$, $\; {\bf n} = {\bf n'}$, and for the moment let us suppose that 
all these integer parameters $\{n_{\alpha}\}$ are {\it different},
$1 \leq n_{1} < n_{2} < ... < n_{M}$. Then, in the summations over the permutations
in Eq.(\ref{C17}), we find two types of terms:

(A) the "diagonal" ones in which the two
permutations coincide, $P = P'$ ;

(B) the "off-diagonal" ones in which
the two permutations are different, $P \not= P'$.

The contribution of the "diagonal" ones reeds
\begin{eqnarray}
\nonumber
Q^{(M,M)^{(A)}}_{{\bf n},{\bf n}}({\bf q}, {\bf q'}) &=&
C^{(M)}_{\bf q, n} C^{(M)^{*}}_{\bf q', n} \;
{\sum_{P}}'
\int_{-\infty}^{+\infty} d^{N}{\bf x}  \; 
\exp\biggl[
-i\sum_{\alpha=1}^{M} q'_{\alpha}\sum_{c\in\Omega_{\alpha}}^{n_{\alpha}} x_{c} 
 -\frac{\kappa}{4}\sum_{\alpha=1}^{M}\sum_{c,c'\in\Omega_{\alpha}}^{n_{\alpha}} |x_{c}-x_{c'}| \biggr] \times
\\
\nonumber
\\
&& \times
\biggl(
\prod_{\substack{ a<b\\ \alpha(a)\not=\alpha(b) }}^{N} \biggl[ 
-\bigl(\partial_{x_a} - \partial_{x_b}\bigr)^{2} + \kappa^{2} \biggr] \biggr) \;
\exp\biggl[
i\sum_{\alpha=1}^{M}q_{\alpha}\sum_{c\in\Omega_{\alpha}}^{n_{\alpha}} x_{c} 
 -\frac{\kappa}{4}\sum_{\alpha=1}^{M}\sum_{c,c'\in\Omega_{\alpha}}^{n_{\alpha}} 
|x_{c}-x_{c'}| \biggr]
\label{C18}
\end{eqnarray}
It is evident that all permutations $\alpha(a)$ in the above equation give the same contribution
and therefore it is sufficient to consider only the contribution of the "trivial"
permutation which is represented by line in Eq.(\ref{C3}). The cluster  ordering
given by this permutation we denote by $\alpha_{0}(a)$. For this particular configuration
of clusters we can redefine the particles numbering, so that instead of a "plane"
index $a = 1, 2, ..., N$ the particles would be counted by two indices $(\alpha, r)$:
$\{x_{a}\} \to \{x_{r}^{\alpha}\} \; (\alpha = 1, ..., M) \; (r = 1, ..., n_{\alpha})$
indicating to which cluster $\alpha$ a given particle belongs and what is
its "internal" cluster number $r$. Due to the symmetry of the integrated expression
in Eq.(\ref{C18}) with respect to the permutations of the particles inside the clusters,
we can introduce the "internal" particles ordering 
for every cluster: $x_{1}^{\alpha}  < x_{2}^{\alpha} < ... < x_{n_{\alpha}}^{\alpha}$. 
In this way, using the relation, Eq.(\ref{C11}), we get
\begin{eqnarray}
\nonumber
&& Q^{(M,M)^{(A)}}_{{\bf n},{\bf n}}({\bf q}, {\bf q'}) \; = \; 
C^{(M)}_{\bf q, n} C^{(M)^{*}}_{\bf q', n} \; \frac{N!}{n_{1}! n_{2}! ... n_{M}!}
\biggl[\prod_{\alpha=1}^{M}
\biggl(
n_{\alpha}!
\int_{-\infty}^{+\infty} dx_{1}^{\alpha}
\int_{x_{1}^{\alpha}}^{+\infty} dx_{2}^{\alpha}
....
\int_{x_{n_{\alpha}-1}^{\alpha}}^{+\infty} dx_{n_{\alpha}}^{\alpha}
\biggr) \biggr] \times
\\
\nonumber
\\
\nonumber
&& \times
\exp\biggl[
-i\sum_{\alpha=1}^{M} q'_{\alpha}\sum_{r=1}^{n_{\alpha}} x_{r}^{\alpha} 
+\frac{\kappa}{2}\sum_{\alpha=1}^{M}\sum_{r=1}^{n_{\alpha}} (n_{\alpha} +1 - 2r)x_{r}^{\alpha} \biggr] \times
\\
\nonumber
\\
&& \times
\biggl(
\prod_{\alpha < \beta}^{M} \prod_{r=1}^{n_{\alpha}} \prod_{r'=1}^{n_{\beta}}
\biggl[ 
-\bigl(\partial_{x_{r}^{\alpha}} - \partial_{x_{r'}^{\beta}}\bigr)^{2} + \kappa^{2} \biggr] \biggr) 
\exp\biggl[
i\sum_{\alpha=1}^{M} q_{\alpha}\sum_{r=1}^{n_{\alpha}} x_{r}^{\alpha} 
+\frac{\kappa}{2}\sum_{\alpha=1}^{M}\sum_{r=1}^{n_{\alpha}} (n_{\alpha} +1 - 2r)x_{r}^{\alpha}\biggr]
\label{C19}
\end{eqnarray}
where the factor $N!/n_{1}!...n_{M}!$ is the total number of permutations of $M$ clusters
over $N$ particles. Taking the derivatives and reorganizing the terms we obtain
\begin{eqnarray}
\nonumber
Q^{(M,M)^{(A)}}_{{\bf n},{\bf n}}({\bf q}, {\bf q'}) &=&
C^{(M)}_{\bf q, n} C^{(M)^{*}}_{\bf q', n} \; N! \; 
\biggl(
\prod_{\alpha < \beta}^{M} \prod_{r=1}^{n_{\alpha}} \prod_{r'=1}^{n_{\beta}}
\bigg| 
\bigl(q_{\alpha} - \frac{i\kappa}{2} n_{\alpha}\bigr) -
\bigl(q_{\beta}  - \frac{i\kappa}{2} n_{\beta} \bigr) +
i\kappa \; (r - r' - 1) \bigg|^{2} \biggr) \times
\\
\nonumber
\\
&& \times
\prod_{\alpha=1}^{M} \Biggr\{
\int_{-\infty}^{+\infty} dx_{1}^{\alpha}
\int_{x_{1}^{\alpha}}^{+\infty} dx_{2}^{\alpha}
....
\int_{x_{n_{\alpha}-1}^{\alpha}}^{+\infty} dx_{n_{\alpha}}^{\alpha} \; 
\mbox{\Large e}^{ \;
i(q_{\alpha}-q'_{\alpha}) \sum_{r=1}^{n_{\alpha}} x_{r}^{\alpha} 
+\kappa\sum_{r=1}^{n_{\alpha}} (n_{\alpha} +1 - 2r)x_{r}^{\alpha}}
\Biggr\}
\label{C20}
\end{eqnarray}
Simple integrations over $x_{r}^{\alpha}$ yields (cf. Eqs.(\ref{B8})-(\ref{B10}))
\begin{eqnarray}
\nonumber
Q^{(M,M)^{(A)}}_{{\bf n},{\bf n}}({\bf q}, {\bf q'}) &=&
\big|C^{(M)}_{\bf q, n}\bigr|^{2}  \; N! \; 
\biggl(
\prod_{\alpha < \beta}^{M} \prod_{r=1}^{n_{\alpha}} \prod_{r'=1}^{n_{\beta}}
\bigg| 
\bigl(q_{\alpha} - \frac{i\kappa}{2} n_{\alpha}\bigr) -
\bigl(q_{\beta}  - \frac{i\kappa}{2} n_{\beta} \bigr) +
i\kappa \; (r - r' - 1) \bigg|^{2} \biggr) \times
\\
\nonumber
\\
&& \times
\prod_{\alpha=1}^{M} \Biggr[
\frac{n_{\alpha}\kappa}{(n_{\alpha}!)^{2} \kappa^{n_{\alpha}}} \; 
(2\pi) \delta(q_{\alpha} - q'_{\alpha}) 
\Biggr]
\label{C21}
\end{eqnarray}

Now let us prove that the  "off-diagonal" terms of Eq.(\ref{C17}),
in which the permutations $P$ and $P'$ are different, give no contribution.
Here we can also chose one of the permutations, say the permutation $P$, to be the "trivial" one 
represented by line in Eq.(\ref{C3}) with the cluster  ordering
denoted by $\alpha_{0}(a)$. Given the symmetry of the wave functions 
it will be sufficient to consider the contribution of the sector
$x_{1} < x_{2} < ... < x_{N}$.
According to Eq.(\ref{C17}), we get
\begin{eqnarray}
\nonumber
Q^{(M,M)^{(B)}}_{{\bf n},{\bf n}}({\bf q}, {\bf q'}) &\propto&
 {\sum_{P'}}' (-1)^{[P']}  
\int_{x_{1} < ... < x_{N}} d^{N}{\bf x}  \; 
\exp\biggl[
-i\sum_{\alpha=1}^{M}q'_{\alpha}\sum_{a\in\Omega'_{\alpha}}^{n_{\alpha}} x_{a} 
 -\frac{\kappa}{4}\sum_{\alpha=1}^{M}\sum_{a,b\in\Omega'_{\alpha}}^{n_{\alpha}} |x_{a}-x_{b}| \biggr] \times
\\
\nonumber
\\
\nonumber
&& \times
\biggl(
\prod_{\substack{ a<b\\ \alpha_{0}(a)\not=\alpha_{0}(b) }}^{N} \biggl[ 
-i\bigl(\partial_{x_a} - \partial_{x_b}\bigr) +i \kappa \sgn(x_{a}-x_{b}) \biggr] \biggr)
\biggl(
\prod_{\substack{ a<b\\ \alpha'(a)\not=\alpha'(b) }}^{N} \biggl[ 
-i\bigl(\partial_{x_a} - \partial_{x_b}\bigr) -i \kappa \sgn(x_{a}-x_{b}) \biggr] \biggr)
\\
\nonumber
\\
&& \times
\exp\biggl[
i\sum_{\alpha=1}^{M}q_{\alpha}\sum_{a\in\Omega^{o}_{\alpha}}^{n_{\alpha}} x_{a} 
 +\frac{\kappa}{2}\sum_{\alpha=1}^{M}\sum_{a\in\Omega^{o}_{\alpha}}^{n_{\alpha}} (n_{\alpha}+1-2r(a))x_{a} \biggr]
\label{C22}
\end{eqnarray}
Here the symbols $\{\Omega^{o}_{\alpha}\}$  denote the clusters 
of the trivial permutation $\alpha_{0}(a)$.
Since $P' \not= P$,
some of the clusters $\Omega'_{\alpha}$ must be different from $\Omega^{o}_{\alpha}$.
As an illustration, let us consider a particular case of $N=10$, with three clusters
$n_{1}=5$ (denoted by the symbol "$\bigcirc$") , $n_{2}=2$ (denoted by the symbol "$\times$")
and $n_{3}=3$ (denoted by the symbol "$\triangle$"):

\vspace{5mm}

\begin{center}

\begin{tabular}{|c||c|c|c|c|c|c|c|c|c|c|}
 \hline
particle number $a$ & 1 & 2 & 3 & 4 & 5 & 6 & 7 & 8 & 9 & 10 \\
\hline
permutation $\alpha_{0}(a)$ &$\bigcirc$&$\bigcirc$&$\bigcirc$&$\bigcirc$&$\bigcirc$&$\times$&$\times$&$\triangle$&$\triangle$&$\triangle$ \\
\hline
permutation $\alpha'(a)$ &$\bigcirc$&$\bigcirc$&$\bigcirc$&$\triangle$&$\bigcirc$&$\times$&$\times$&$\bigcirc$&$\triangle$&$\triangle$ \\
\hline
\end{tabular}

\end{center}

\vspace{5mm}
Here in the permutation $\alpha'(a)$ the particle $a=4$ belong to the cluster $\alpha=3$ 
(and not to the cluster $\alpha=1$ as in the permutation $\alpha_{0}(a)$), and the 
particle $a=8$ belong to the cluster $\alpha=1$ (and not to the cluster $\alpha=3$
as in the permutation $\alpha_{0}(a)$). Now let us look carefully at the structure of the 
products in Eq.(\ref{C22}). Unlike the first product, which contains no "internal"
products among particles belonging to the cluster $\Omega^{o}_{1}$, the second product
{\it does}. Besides, the signs of the differential operators
$\bigl(\partial_{x_a} - \partial_{x_b}\bigr)$ in the second product 
is {\it opposite} to the "normal" ones in the first product (cf. Eqs.(\ref{C7})-(\ref{C9})).
It is these two factors (the presence of the "internal" products and the "wrong" signs
of the differential operators) which makes the "off-diagonal" contributions, Eq.(\ref{C22}),
to be zero. Indeed, in the above example, the second product contains the term
\begin{equation}
 \label{C23}
\Pi'_{4,5} \; \equiv \; 
\biggl[-i\bigl(\partial_{x_4} - \partial_{x_5}\bigr) + i \kappa  \biggr] 
\exp\biggl[
i\sum_{\alpha=1}^{3} q_{\alpha}\sum_{a\in\Omega^{o}_{\alpha}}^{n_{\alpha}} x_{a} 
 +\frac{\kappa}{2}\sum_{\alpha=1}^{3}\sum_{a\in\Omega^{o}_{\alpha}}^{n_{\alpha}} (n_{\alpha}+1-2r(a))x_{a} \biggr]
\end{equation}
(we remind that the particles in the clusters $\Omega^{o}_{\alpha}$ are ordered,
and in particular  $x_{4} < x_{5}$).
Taking the derivatives, we get
\begin{eqnarray}
 \nonumber
\Pi'_{4,5}  &\propto&  
\biggl[-\biggl(iq_{1} + \frac{\kappa}{2}\bigl(n_{1}+1-2r(4)\bigr) 
             - iq_{1} - \frac{\kappa}{2}\bigl(n_{1}+1-2r(5)\bigr) \biggr) + \kappa \Biggr] \\
&\propto&
\bigl[ r(4)-r(5) + 1 \bigr] \; = \; 0
\label{C24}
\end{eqnarray}
since in the first cluster $r(a)=a$. 

One can easily understand that the above example reflect 
the general situation. Since all the cluster sizes $n_{\alpha}$ are supposed to be different,
whatever the permutation $\alpha'(a)$ is, we can always find a cluster
$\Omega^{o}_{\alpha}$ 
such that some of its particles belong to the same cluster number $\alpha$ in the permutation
$\alpha'(a)$ while the others do not. Then one has to consider the contribution of
the product of two {\it neighboring number} points
\begin{equation}
 \label{C25}
\Pi'_{k, k+1} \; = \; 
\biggl[-i\bigl(\partial_{x_k} - \partial_{x_{k+1}}\bigr) + i \kappa  \biggr] 
\exp\biggl[
i\sum_{\alpha=1}^{M} q_{\alpha}\sum_{a\in\Omega^{o}_{\alpha}}^{n_{\alpha}} x_{a} 
 +\frac{\kappa}{2}\sum_{\alpha=1}^{M}\sum_{a\in\Omega^{o}_{\alpha}}^{n_{\alpha}} (n_{\alpha}+1-2r(a))x_{a} \biggr]
\end{equation}
where in the permutation $\alpha'(a)$ the particle $k$ belong to the cluster number $\alpha$ 
and the particle $(k+1)$ belong to some
other cluster. Taking the derivatives
one gets
\begin{equation}
 \label{C26}
\Pi'_{k, k+1} \; \propto \; 
\bigl[r(k)-r(k+1) + 1\bigr] \; = \; 0
\end{equation}
as $r(a)$ is the ''internal`` particle number in the cluster $\Omega^{o}_{\alpha}$,
where $r(k+1) = r(k) + 1$ (cf. Eqs.(\ref{C7})-(\ref{C9})).

Thus, the only non-zero contribution to the overlap, Eq.(\ref{C15}), of two wave function 
$\Psi_{\bf q', n}^{(M)}({\bf x})$ and $\Psi_{\bf q, n}^{(M)}({\bf x})$ (having the same
number of clusters $M$ and characterized by
the same set of the integer parameters  $1 \leq n_{1} < n_{2} < ... < n_{M}$) comes from the "diagonal" terms, Eq.(\ref{C21}):
\begin{eqnarray}
\nonumber
Q^{(M,M)}_{{\bf n},{\bf n}}({\bf q}, {\bf q'}) &=&
\big|C^{(M)}_{\bf q, n}\big|^{2} \; N!  
\prod_{\alpha=1}^{M} \Biggr[
\frac{n_{\alpha}\kappa}{(n_{\alpha}!)^{2} \kappa^{n_{\alpha}}} 
\Biggr] 
\biggl(
\prod_{\alpha < \beta}^{M} \prod_{r=1}^{n_{\alpha}} \prod_{r'=1}^{n_{\beta}}
\bigg| 
\bigl(q_{\alpha} - \frac{i\kappa}{2} n_{\alpha} \bigr) -
\bigl(q_{\beta}  - \frac{i\kappa}{2} n_{\beta}  \bigr)
+i\kappa \; (r - r' - 1)
\bigg|^{2} \biggr) 
\times
\\
\nonumber
\\
&& \times
\prod_{\alpha=1}^{M} \Biggr[
(2\pi) \delta(q_{\alpha} - q'_{\alpha}) 
\Biggr]
\label{C27}
\end{eqnarray}

The situation when there are clusters which have the same numbers of particles $n_{\alpha}$ 
is somewhat more complicated. Let us consider the overlap between two wave function
$\Psi_{\bf q', n}^{(M)}({\bf x})$ and $\Psi_{\bf q, n}^{(M)}({\bf x})$ (which, as before
have the same $M$ and ${\bf n}$) such that in the set of $M$ integers
$n_{1}, n_{2}, ... , n_{M}$ there are two $n_{\alpha}$'s which are equal, say
$n_{\alpha_{1}} = n_{\alpha_{2}}$ (where $\alpha_{1} \not= \alpha_{2}$).
In the eigenstate $({\bf q', n})$ these two clusters have 
the center of mass momenta $q'_{\alpha_{1}}$ and $q'_{\alpha_{2}}$, and in the 
the eigenstate $({\bf q, n})$ they have the momenta $q_{\alpha_{1}}$ and $q_{\alpha_{2}}$
correspondingly.
According to the above discussion, the non-zero contributions in the summation
over the cluster permutations $\alpha(a)$ and $\alpha'(a)$ in Eq.(\ref{C17})
appears only if the clusters $\{\Omega_{\alpha}\}$ of the permutation $\alpha(a)$
totally coincide with the clusters $\{\Omega'_{\alpha}\}$ of the permutation $\alpha'(a)$.
In the case when all $n_{\alpha}$ are different this is possible only if 
the permutation $\alpha(a)$ coincides with the permutation $\alpha'(a)$. In contrast to that,
in the case when we have $n_{\alpha_{1}} = n_{\alpha_{2}}$, there are {\it two}
non-zero options. The first one, as before, is given by the "diagonal" terms
with $\alpha(a) = \alpha'(a)$ (so that the clusters $\{\Omega_{\alpha}\}$ and
$\{\Omega'_{\alpha}\}$ are just the same), and this contribution
is proportional to $\delta(q_{\alpha_{1}} - q'_{\alpha_{1}}) \, \delta(q_{\alpha_{2}} - q'_{\alpha_{2}})$.
The second  ("off-diagonal") contribution is given by such permutation $\alpha'(a)$
in which the cluster $\Omega'_{\alpha_{1}}$ (of the permutation $\alpha'(a)$) coincide
with the cluster $\Omega_{\alpha_{2}}$ (of the permutation $\alpha(a)$) and 
the cluster $\Omega'_{\alpha_{2}}$ (of the permutation $\alpha'(a)$) coincide
with the cluster $\Omega_{\alpha_{1}}$ (of the permutation $\alpha(a)$) while the 
rest of the clusters of these two permutations are the same, 
$\Omega'_{\alpha} = \Omega_{\alpha} \; (\alpha \not= \alpha_{1}, \alpha_{2})$.
Correspondingly, this last contribution is proportional to 
$\delta(q_{\alpha_{1}} - q'_{\alpha_{2}}) \, \delta(q_{\alpha_{2}} - q'_{\alpha_{1}})\; (-1)^{n_{\alpha_{1}}}$.
In fact this situation with two equivalent contributions is the consequence of the symmetry
of the wave function $\Psi_{\bf q', n}^{(M)}({\bf x})$: 
the permutation of two momenta $q_{\alpha_{1}}$ and $q_{\alpha_{2}}$ belonging to 
the clusters which have the same numbers of particles, 
$n_{\alpha_{1}}=n_{\alpha_{2}}$ produces the factor $(-1)^{n_{\alpha_{1}}}$.
This is evident from the general expression for the wave function, eq.(\ref{A9}),
where the permutation of any two momenta $q_{\alpha_{1}}$ and $q_{\alpha_{2}}$ belonging to 
the clusters which have the same numbers of particles corresponds to the permutation of
$n$ columns of the matrix $\exp(i q_{a} x_{b})$. Therefore considering the clusters with equal numbers of particles as equivalent and restricting analysis to the sectors 
$q_{\alpha_{1}} < q_{\alpha_{2}}; \; q'_{\alpha_{1}} < q'_{\alpha_{2}}$
we find that the second contribution,  
$\delta(q_{\alpha_{1}} - q'_{\alpha_{2}}) \, \delta(q_{\alpha_{2}} - q'_{\alpha_{1}})$
is identically equal to zero, thus returning to the above result Eq.(\ref{C27}).

A generic eigenstate
$({\bf q}, {\bf n})$ with $M$ clusters could be specified in terms of the 
following set of parameters:
\begin{equation}
\label{C28}
({\bf q}, {\bf n}) \; = \;  \{
(\underbrace{q_{1}, m_{1}), ..., (q_{s_{1}}, m_{1})}_{s_{1}}; 
 \underbrace{(q_{s_{1}+1}, m_{2}), ..., (q_{s_{1}+s_{2}}, m_{2})}_{s_{2}}; 
\; .... \; ; 
 \underbrace{(q_{s_{1}+...+s_{k-1}+1}, m_{k}), ..., (q_{s_{1}+...+s_{k}}, m_{k})}_{s_{k}}\}
\end{equation}
where  $s_{1}+s_{2}+ ... +s_{k} = M$ and $k$ integers $\{ m_{i}\}$ ($ 1\leq k \leq M$)
are all supposed to be {\it different}: 
\begin{equation}
 \label{C29}
1 \leq m_{1} < m_{2} < ... < m_{k}
\end{equation}
Here the integer parameter $k$ denotes the number of different cluster types. For a given $k$
\begin{equation}
\label{C30}
\sum_{\alpha=1}^{M} n_{\alpha} \; = \; \sum_{i=1}^{k} s_{i} m_{i} \; =  \; N
\end{equation}
Due to the symmetry  with respect to the momenta permutations inside the subsets 
of equal $n$'s it is sufficient to consider the wave functions in the sectors
\begin{eqnarray}
\label{C31}
&&
q_{1} < q_{2} < ... < q_{s_{1}} \; ;\\
\nonumber
&&
q_{s_{1}+1} < q_{s_{1}+2} < ... < q_{s_{1}+s_{2}} \; ; \\
\nonumber
&&
................. \\
\nonumber
&&
q_{s_{1}+...+s_{k-1}+1} < q_{s_{1}+...+s_{k-1}+2} < ... < q_{s_{1}+...+s_{k-1}+s_{k}}
\end{eqnarray}
In this representation we again recover the above result Eq.(\ref{C27})

Finally, let us consider the overlap of two eigenstates described by two {\it different}
sets of integer parameters,  ${\bf n'} \not= {\bf n}$. In fact this situation is quite simple
because if the clusters of the two states are different from each other, it means
that in the summation over the pairs of permutations $P$ and $P'$ in Eq.(\ref{C17})
there exist no two permutations for which these two sets of clusters $\{\Omega_{\alpha}\}$ and 
$\{\Omega'_{\alpha}\}$ would coincide. Which, according to the above analysis, means that 
this expression is equal to zero. Note that the condition $M' \not= M$ automatically implies 
that ${\bf n'} \not= {\bf n}$. 

Thus we have proved that
\begin{eqnarray}
\nonumber
Q^{(M,M')}_{{\bf n},{\bf n'}}({\bf q}, {\bf q'}) &=& 
\big|C^{(M)}_{\bf q, n}\big|^{2}  \;
{\boldsymbol \delta}(M,M') \; \biggl(\prod_{\alpha=1}^{M} 
{\boldsymbol \delta}(n_{\alpha},n'_{\alpha}) \biggr)
\biggl(\prod_{\alpha=1}^{M} (2\pi) \delta(q_{\alpha}-q'_{\alpha}) \biggr)
\times
\\
\nonumber
\\
&& \times
 N! \; 
\prod_{\alpha=1}^{M} \Biggr[
\frac{n_{\alpha}\kappa}{(n_{\alpha}!)^{2} \kappa^{n_{\alpha}}} 
\Biggr]
\prod_{\alpha < \beta}^{M} \prod_{r=1}^{n_{\alpha}} \prod_{r'=1}^{n_{\beta}}
\Big| 
\bigl(q_{\alpha} - \frac{i\kappa}{2} n_{\alpha} \bigr) -
\bigl(q_{\beta}  - \frac{i\kappa}{2} n_{\beta}  \bigr)
+i\kappa \; (r - r' - 1)
\Big|^{2}  
\label{C32}
\end{eqnarray}
where the integer parameters $\{n_{\alpha}\}$ and $\{n'_{\alpha}\}$ are assumed to have the 
generic structure represented in Eqs.(\ref{C28})-(\ref{C30}), and the  momenta 
$\{q_{\alpha}\}$ and $\{q'_{\alpha}\}$ of the clusters with equal numbers of particles 
are restricted in the sectors, Eq.(\ref{C31}). 
According to Eq.(\ref{C32}), the orthonormality condition defines the normalization
constant 
\begin{equation}
   \label{C33}
  \big| C^{(M)}({\bf q,n})\big|^{2} = 
\frac{1}{N!} \; 
\biggl[
\prod_{\alpha=1}^{M} \frac{(n_{\alpha}!)^{2}\kappa^{n_{\alpha}} }
   {n_{\alpha} \kappa}
\biggr]
\prod_{\alpha < \beta}^{M} \prod_{r=1}^{n_{\alpha}} \prod_{r'=1}^{n_{\beta}}
\frac{1}{\Big|\bigl(q_{\alpha} - \frac{i\kappa}{2} n_{\alpha} \bigr) -
\bigl(q_{\beta}  - \frac{i\kappa}{2} n_{\beta}  \bigr)
+i\kappa \; (r - r' - 1)\Big|^{2}}
\end{equation}
In other words, the wave functions, Eq.(\ref{C14}), 
form the orthonormal set. Although, at present we are not able to prove that
this set is complete, the {\it suggestion} of the completeness 
(which assumes that there are exist no other eigenstates besides those described above)
looks quite natural.

\vspace{10mm}

{\center {\bf 4. Propagator}}

\vspace{3mm}

The time dependent solution $ \Psi({\bf x},t)$ of the imaginary-time 
Schr\"odinger equation
\begin{equation}
   \label{C34}
\beta \partial_t \Psi({\bf x}; t) \; = \;
\frac{1}{2}\sum_{a=1}^{N}\partial_{x_a}^2 \Psi({\bf x}; t)
  \; + \; \frac{1}{2}\kappa \sum_{a\not=b}^{N} \delta(x_a-x_b) \Psi({\bf x}; t)
\end{equation}
with the initial condition 
\begin{equation}
   \label{C35}
\Psi({\bf x}; 0) = \Pi_{a=1}^{N} \delta(x_a)
\end{equation}
 can be represented in terms of the 
linear combination of the eigenfunctions $\Psi_{\bf q, n}^{(M)}({\bf x})$, 
Eq.(\ref{C14}):
\begin{equation}
\label{C36}
\Psi({\bf x},t) \; = \; \sum_{M=1}^{N} \;
{\sum_{{\bf n}}}' \; 
\int ' {\cal D} {\bf q} \; \; 
 \Psi_{\bf q, n}^{(M)}({\bf x}) \Psi_{\bf q, n}^{(M)^{*}}({\bf 0}) \; 
\exp\bigl[-E_{M}({\bf q,n}) \; t \bigr]
\end{equation}
where the energy spectrum $E_{M}({\bf q,n})$ is given by Eq.(\ref{C14a}).  
The summations over $n_{\alpha}$ are performed here in terms of the parameters $\{s_{i}, m_{i}\}$,
Eqs.(\ref{C28})-(\ref{C30}):
\begin{equation}
\label{C37}
{\sum_{{\bf n}}}' \; \equiv \;
\sum_{k=1}^{M} \; \; 
\sum_{s_{1}...s_{k}=1}^{\infty} \; \; 
\sum_{1 \leq m_{1} ... < m_{k}}^{\infty}
\; {\boldsymbol \delta}\biggl(\sum_{i=1}^{k} s_{i}, \; M\biggr)
\; \;{\boldsymbol \delta}\biggl(\sum_{i=1}^{k} s_{i} m_{i}, \; N\biggr)
\end{equation}
where ${\boldsymbol \delta}(n,l)$ is the Kronecker symbol,
and for simplicity (due to the presence of these Kronecker symbols)
the summations over $m_{i}$ and $s_{i}$
are extended to infinity. The symbol $\int '{\cal D} {\bf q}$ in Eq.(\ref{C36}) 
denotes the integration 
over $M$ momenta $q_{\alpha}$ in the sectors, Eq.(\ref{C31}). 

The replica partition function $Z(N,L)$ of the original directed polymer problem
is obtained via a particular choice of the final-point coordinates,
\begin{equation}
\label{C38}
Z(N,L) \; = \; \Psi({\bf 0};L) \; = \; 
\sum_{M=1}^{N} \;
{\sum_{{\bf n}}}' \; 
\int ' {\cal D} {\bf q} \; \; 
\bigl| \Psi_{\bf q, n}^{(M)}({\bf 0}) \bigr|^2 \; 
\exp\bigl[-E_{M}({\bf q,n}) \; L \bigr]
\end{equation}
According to Eq.(\ref{C14}), for $M \geq 2$,
\begin{eqnarray}
\nonumber
\Psi_{\bf q, n}^{(M)}({\bf 0}) 
&=& 
C^{(M)}_{\bf q, n}
{\sum_{P}}' (-1)^{[P]} 
\prod_{\substack{ a<b\\ \alpha(a)\not=\alpha(b) }}^{N}
\biggl[ 
  \Bigl(q_{\alpha(a)} - \frac{i\kappa}{2} \big[n_{\alpha(a)} + 1 - 2r(a)\big] \Bigr)
- \Bigl(q_{\alpha(b)} - \frac{i\kappa}{2} \big[n_{\alpha(b)} + 1 - 2r(b)\big] \Bigr)
\biggr] 
\\
\nonumber
\\
&=& 
C^{(M)}_{\bf q, n} \; \frac{N!}{n_{1}! n_{2}! ... n_{M}!} \;
\prod_{\alpha<\beta}^{M} \prod_{r=1}^{n_{\alpha}} \prod_{r'=1}^{n_{\beta}}
\biggl[ 
\Big(q_{\alpha} - \frac{i\kappa}{2} n_{\alpha} \Big) 
- \Big(q_{\beta} - \frac{i\kappa}{2} n_{\beta} \Big) 
+ i \kappa \bigl(r - r'\bigr)
\biggr]
\label{C39}
\end{eqnarray}
Substituting here the value of the normalization constant, Eq.(\ref{C33}), we get
\begin{equation}
 \label{C40}
\bigl| \Psi_{\bf q, n}^{(M)}({\bf 0}) \bigr|^2 \; = \; 
\frac{N! \kappa^{N}}{ \Bigl(\prod_{\alpha=1}^{M} \kappa n_{\alpha}\Bigr)} \; 
\prod_{\alpha<\beta}^{M} 
\frac{
\prod_{r=1}^{n_{\alpha}} \prod_{r'=1}^{n_{\beta}}
\Big|
\bigl(q_{\alpha} - \frac{i\kappa}{2} n_{\alpha} \bigr) -
\bigl(q_{\beta}  - \frac{i\kappa}{2} n_{\beta}  \bigr)
+i\kappa \; (r - r')
\Big|^{2}}{
\prod_{r=1}^{n_{\alpha}} \prod_{r'=1}^{n_{\beta}}
\Big|
\bigl(q_{\alpha} - \frac{i\kappa}{2} n_{\alpha} \bigr) -
\bigl(q_{\beta}  - \frac{i\kappa}{2} n_{\beta}  \bigr)
+i\kappa \; (r - r'-1)
\Big|^{2}}
\end{equation}
This expression can be essentially simplified. 
Shifting the product over $r'$ in the denominator by $1$ we obtain
\begin{equation}
 \label{C41}
\bigl| \Psi_{\bf q, n}^{(M)}({\bf 0}) \bigr|^2 \; = \; 
\frac{N! \kappa^{N}}{ \Bigl(\prod_{\alpha=1}^{M} \kappa n_{\alpha}\Bigr)} \; 
\prod_{\alpha<\beta}^{M} 
\frac{
\prod_{r=1}^{n_{\alpha}} 
\Big|
\bigl(q_{\alpha} - \frac{i\kappa}{2} n_{\alpha} \bigr) -
\bigl(q_{\beta}  - \frac{i\kappa}{2} n_{\beta}  \bigr)
+i\kappa \; (r - 1)
\Big|^{2}}{
\prod_{r=1}^{n_{\alpha}}
\Big|
\bigl(q_{\alpha} - \frac{i\kappa}{2} n_{\alpha} \bigr) -
\bigl(q_{\beta}  - \frac{i\kappa}{2} n_{\beta}  \bigr)
+i\kappa \; (r - n_{\beta} -1)
\Big|^{2}}
\end{equation}
Redefining the product parameter $r$ in the denominator, $ r \to n_{\alpha} + 1 - r$, 
and changing the obtained expression (under the modulus square) by its complex conjugate
we get
\begin{equation}
 \label{C42}
\bigl| \Psi_{\bf q, n}^{(M)}({\bf 0}) \bigr|^2 \; = \; 
\frac{N! \kappa^{N}}{ \Bigl(\prod_{\alpha=1}^{M} \kappa n_{\alpha}\Bigr)} \; 
\prod_{\alpha<\beta}^{M} 
\frac{
\prod_{r=1}^{n_{\alpha}} 
\Big|
\bigl(q_{\alpha} - \frac{i\kappa}{2} n_{\alpha} \bigr) -
\bigl(q_{\beta}  - \frac{i\kappa}{2} n_{\beta}  \bigr)
+i\kappa \; (r - 1)
\Big|^{2}}{
\prod_{r=1}^{n_{\alpha}}
\Big|
\bigl(q_{\alpha} - \frac{i\kappa}{2} n_{\alpha} \bigr) -
\bigl(q_{\beta}  - \frac{i\kappa}{2} n_{\beta}  \bigr)
+i\kappa \; r
\Big|^{2}}
\end{equation}
Shifting now the product over $r$ in the numerator by $1$ we finally obtain
\begin{equation}
 \label{C43}
\bigl| \Psi_{\bf q, n}^{(M)}({\bf 0}) \bigr|^2 \; = \; 
\frac{N! \kappa^{N}}{ \Bigl(\prod_{\alpha=1}^{M} \kappa n_{\alpha}\Bigr)} \; 
\prod_{\alpha<\beta}^{M} 
\frac{\big|q_{\alpha}-q_{\beta} -\frac{i\kappa}{2}(n_{\alpha}-n_{\beta})\big|^{2}}{
      \big|q_{\alpha}-q_{\beta} -\frac{i\kappa}{2}(n_{\alpha}+n_{\beta})\big|^{2}}
\end{equation}
For $M=1$, according to Eqs.(\ref{B1}) and (\ref{B11}),
\begin{equation}
\label{C44}
\big|\Psi_{q}^{(1)}({\bf 0})\big|^{2} \; = \; \frac{\kappa^N N!}{\kappa N}
\end{equation}
Since the function $f\bigl({\bf q}, {\bf n}\bigr) = \bigl| \Psi_{\bf q, n}^{(M)}({\bf 0}) \bigr|^2 \; 
\exp\bigl[-E_{M}({\bf q,n}) \; L \bigr]$ in Eq.(\ref{C38}) is symmetric 
with respect to permutations of all its $M$ pairs of arguments $(q_{\alpha}, n_{\alpha})$
the integrations over $M$ momenta $q_{\alpha}$ can be extended beyond the sector defined in Eq.(\ref{C31})
for the whole space $R_{M}$. As a consequence, there is no need 
to distinguish equal and different $n_{\alpha}$'s any more, and instead of Eq.(\ref{C37}), we can sum
over $M$ integer parameters $n_{\alpha}$ with the only constrain, Eq.(\ref{C2})
(note that this kind of simplifications holds only for the specific "zero final-point" object
$\Psi({\bf 0};t)$, Eq.(\ref{C38}), and {\it not} for the general propagator
$\Psi({\bf x};t)$, Eq.(\ref{C36}) containing $N$ arbitrary coordinates $x_{1}, ..., x_{N}$).
Thus, instead of Eq.(\ref{C38}) we get
\begin{equation}
\label{C45}
Z(N,L) \; = \;  
\sum_{M=1}^{N} \; 
\frac{1}{M!}
\biggl[
\prod_{\alpha=1}^{M} 
\int_{-\infty}^{+\infty} \; \frac{dq_{\alpha}}{2\pi} 
\sum_{n_{\alpha}=1}^{\infty}
\biggr] \;
{\boldsymbol \delta}\biggl(\sum_{\alpha=1}^{M} n_{\alpha}, \; N \biggr) \;
\big| \Psi_{\bf q, n}^{(M)}({\bf 0}) \big|^2 \; 
\mbox{\LARGE e}^{-E_{M}({\bf q,n}) L}
\end{equation}
Substituting here Eqs.(\ref{C14a}), (\ref{C43}) and (\ref{C44}) we get the following sufficiently
compact representation for the replica partition function:
\begin{eqnarray}
\nonumber
Z(N.L) &=& 
N! \; \kappa^{N} \biggl\{
 \int_{-\infty}^{+\infty} \frac{dq}{2\pi\kappa N} \;
\exp\Bigl[
-\frac{L}{2\beta} N q^{2} + \frac{\kappa^{2}L}{24\beta} (N^{3} -N) 
\Bigr] \; +
\\
\nonumber
\\
\nonumber
&+& 
\sum_{M=2}^{N} \frac{1}{M!} 
\biggl[
\prod_{\alpha=1}^{M}
\sum_{n_{\alpha}=1}^{\infty}
\int_{-\infty}^{+\infty} \frac{d q_{\alpha}}{2\pi\kappa n_{\alpha}} 
\biggr]
\;{\boldsymbol \delta}\biggl(\sum_{\alpha=1}^{M} n_{\alpha}, \; N \biggr) 
\prod_{\alpha<\beta}^{M} 
\frac{\big|q_{\alpha}-q_{\beta} -\frac{i\kappa}{2}(n_{\alpha}-n_{\beta})\big|^{2}}{
      \big|q_{\alpha}-q_{\beta} -\frac{i\kappa}{2}(n_{\alpha}+n_{\beta})\big|^{2}} 
\times
\\
&\times&
\exp\Bigl[
-\frac{L}{2\beta}\sum_{\alpha=1}^{M} n_{\alpha} q_{\alpha}^{2} + 
\frac{\kappa^{2}L}{24\beta} \sum_{\alpha=1}^{M} (n_{\alpha}^{3} - n_{\alpha})
\Bigr] 
\biggr\}
\label{C46}
\end{eqnarray}
The first term in the above expression is the contribution of the ground state $(M=1)$,
while  the next terms $(M \geq 2)$ are the contributions of the rest of the energy
spectrum. After simple algebra the above replica partition function can be represented as follows:
\begin{equation}
   \label{C47}
 Z(N,L) \; =  \mbox{\LARGE e}^{-\beta N L f_{0}} \;\;
\tilde{Z}(N,\lambda) 
\end{equation}
where $f_{0} = \frac{1}{24}\beta^4 u^2 - \frac{1}{\beta L} \ln(\beta^{3} u)$, 
and
\begin{eqnarray}
\nonumber
\tilde{Z}(N.L) &=& 
N! \; \biggl\{
 \int_{-\infty}^{+\infty} \frac{dq}{2\pi\kappa N} \;
\exp\Bigl[
-\frac{L}{2\beta} N q^{2} + \frac{\kappa^{2}L}{24\beta} N^{3} 
\Bigr]
\; +
\\
\nonumber
\\
\nonumber
&+& 
\sum_{M=2}^{N} \frac{1}{M!} 
\biggl[
\prod_{\alpha=1}^{M}
\sum_{n_{\alpha}=1}^{\infty}
\int_{-\infty}^{+\infty} \frac{d q_{\alpha}}{2\pi\kappa n_{\alpha}} 
\biggr]
\;{\boldsymbol \delta}\biggl(\sum_{\alpha=1}^{M} n_{\alpha}, \; N \biggr) 
\prod_{\alpha<\beta}^{M} 
\frac{\big|q_{\alpha}-q_{\beta} -\frac{i\kappa}{2}(n_{\alpha}-n_{\beta})\big|^{2}}{
      \big|q_{\alpha}-q_{\beta} -\frac{i\kappa}{2}(n_{\alpha}+n_{\beta})\big|^{2}} 
\times
\\
&\times&
\exp\Bigl[
-\frac{L}{2\beta}\sum_{\alpha=1}^{M} n_{\alpha} q_{\alpha}^{2} + 
\frac{\kappa^{2}L}{24\beta} \sum_{\alpha=1}^{M} n_{\alpha}^{3}
\Bigr] 
\biggr\}
\label{C48}
\end{eqnarray}

\vspace{10mm}

\begin{center}

\appendix{\Large Appendix C}

\vspace{5mm}

 {\bf \large Fredholm determinant with the Airy kernel and the Tracy-Widom distribution}

\end{center}

\newcounter{C}
\setcounter{equation}{0}
\renewcommand{\theequation}{C.\arabic{equation}}

\vspace{5mm}

In this Appendix the original derivation of  Tracy and Widom \cite{Tracy-Widom} will be repeated 
in simple terms to demonstarate that the function $F_{2}(s)$
defined as the Fredholm determinant with the Airy kernel can be expressed in terms of the 
solution of the Panlev\'e II differential equation, namely
\begin{equation}
 \label{c1}
 F_{2}(s) \equiv \det\bigl[1 - \hat{K}_{A}\bigr] \; = \; 
\exp\biggl[
-\int_{s}^{\infty} dt \, (t-s) q^{2}(t) 
\biggr]
\end{equation}
where $\hat{K}_{A}$ is the integral operator defined on semi-infinite interval $[s, \infty)$ with the 
Airy kernel, 
\begin{equation}
 \label{c2}
K_{A}(t_{1},t_{2}) \; = \; 
\frac{\Ai(t_{1}) \Ai'(t_{2}) - \Ai'(t_{1}) \Ai(t_{2})}{t_{1} - t_{2}}
\end{equation}
and the function $q(t)$ is the solution of the Panlev\'e II equation,
\begin{equation}
 \label{c3}
q'' = t q + 2 q^{3}
\end{equation}
with the boundary condition, $q(t\to +\infty) \sim \Ai(t)$.

\vspace{5mm}

Let us introduce a new function $R(t)$ such that
\begin{equation}
 \label{c4}
 F_{2}(s) \; = \; 
\exp\biggl[
-\int_{s}^{\infty} dt R(t) 
\biggr]
\end{equation}
or, according to the definition, Eq.(\ref{c1}),
\begin{equation}
 \label{c5}
R(s) \; = \; \frac{d}{ds} \ln\Bigl[\det\bigl(1 - \hat{K}_{A}\bigr) \Bigr]
\end{equation}
Here the logarithm of the determinant can be expressed in terms of the trace:
\begin{eqnarray}
 \label{c6}
\ln\Bigl[\det\bigl(1 - \hat{K}_{A}\bigr) \Bigr] 
&=&
- \sum_{n=1}^{\infty} \frac{1}{n} \, Tr \, \hat{K}_{A}^{n} 
\\
\nonumber
\\
\nonumber
&\equiv& 
- \sum_{n=1}^{\infty} \frac{1}{n}
\int_{s}^{\infty} dt_{1} \int_{s}^{\infty} dt_{2} \, ... \, \int_{s}^{\infty} dt_{n} \; 
 K_{A}(t_{1},t_{2}) K_{A}(t_{2},t_{3}) \, ... \, K_{A}(t_{n},t_{1}) 
\end{eqnarray}
Taking derivative of this expression we gets
\begin{eqnarray}
 \label{c7}
R(s) &=& - \int_{s}^{\infty} dt \, \bigl(1 - \hat{K}_{A}\bigr)^{-1} (s,t) \, K_{A}(t,s)
\\
\nonumber
\\
\nonumber
&\equiv& 
 - K_{A}(s,s) 
- \sum_{n=2}^{\infty} 
\int_{s}^{\infty} dt_{1} \int_{s}^{\infty} dt_{2} \, ... \, \int_{s}^{\infty} dt_{n-1} \; 
K_{A}(s,t_{1}) K_{A}(t_{1},t_{2}) \, ... \, K_{A}(t_{n-1},s) 
\end{eqnarray}
Substituting here the integral representation of the Airy kernel, Eq.(\ref{c2}),
\begin{equation}
 \label{c8}
K_{A}(t_{1},t_{2}) \; = \; 
\int_{0}^{\infty} dz \Ai(t_{1} + z) \, \Ai(t_{2} + z)
\end{equation}
after some efforts in simple algebra one gets
\begin{equation}
 \label{c9}
R(s) \; = \; 
\int_{s}^{\infty} dt_{1} \int_{s}^{\infty} dt_{2} \, 
\Ai(t_{1}) \, \bigl(1 - \hat{K}_{A}\bigr)^{-1} (t_{1},t_{2}) \, \Ai(t_{2})
\end{equation}
Taking the derivative of this expression and applying some more efforts in 
 slightly more complicated algebra, we obtain
\begin{equation}
 \label{c10}
\frac{d}{ds} R(s) \; = \; - q^{2}(s)
\end{equation}
where
\begin{equation}
 \label{c11}
q(s) \; = \; \int_{s}^{\infty} dt \, \bigl(1 - \hat{K}_{A}\bigr)^{-1} (s,t) \, \Ai(t)
\end{equation}
According to Eq.(\ref{c10}),
\begin{equation}
 \label{c10a}
 R(s) \; = \; \int_{s}^{\infty} dt \, q^{2}(t)
\end{equation}
Let us introduce two more functions
\begin{eqnarray}
 \label{c12}
v(s) &=& \int_{s}^{\infty} dt_{1} \int_{s}^{\infty} dt_{2} \, 
\Ai(t_{1}) \, \bigl(1 - \hat{K}_{A}\bigr)^{-1} (t_{1},t_{2}) \, \Ai'(t_{2})
\\
\nonumber
\\
\label{c13}
p(s) &=& \int_{s}^{\infty} dt \, \bigl(1 - \hat{K}_{A}\bigr)^{-1} (s,t) \, \Ai'(t)
\end{eqnarray}
Taking derivatives of the above three functions $q(s)$, $v(s)$ and $p(s)$, 
Eqs.(\ref{c11}), (\ref{c12}) and (\ref{c13}),
after somewhat painfull algebra one finds the following three relations:
 \begin{eqnarray}
 \label{c14}
q' &=& p \; - \; R \, q
\\
\label{c15}
p' &=& s \, q \; - \; p \, R \; - \; 2 q \, v
\\
\label{c16}
v' &=& - p \, q
\end{eqnarray}
Taking derivative of the combination $\bigl(R^{2} - 2v\bigr)$ and using Eqs.(\ref{c10}) and (\ref{c16}),
we get
\begin{equation}
 \label{c17}
\frac{d}{ds} \bigl(R^{2} - 2v\bigr) \; = \; 2q \, (p \; - \; R\, q)
\end{equation}
On the other hand, multiplying Eq.(\ref{c14}) by $2q$ we find
\begin{equation}
 \label{c18}
\frac{d}{ds} q^{2} \; = \; 2q \, (p \; - \; R\, q)
\end{equation}
Comparing Eqs.(\ref{c17}) and (\ref{c18}) and taking into account that 
the value of all the above functions at $s\to\infty$ is zero, we obtain the following relation
\begin{equation}
 \label{c19}
R^{2} - 2v \; = \; q^{2}
\end{equation}
Finally, taking the derivatiove of Eq.(\ref{c14}) and using Eqs.(\ref{c10}), (\ref{c14}), (\ref{c15})
and (\ref{c19}) we easily find
\begin{equation}
 \label{c20}
q'' \; = \; 2q^{3} \; + \; s q
\end{equation}
which is the special case of the Panlev\'e II differential equation \cite{Panleve, Iwasaki}.
Thus, substituting Eq.(\ref{c10a}) into Eq.(\ref{c4}) we obtain Eq.(\ref{c1}).

\vspace{5mm}

The function $F_{2}(s)$ gives the probability 
that a random quantity $t$ described by a probability distribution functions
$P_{TW}(t)$ 
has the value bigger than a given parameter $s$:
\begin{equation}
 \label{c21}
F_{2}(s) \; = \; \int_{s}^{\infty} dt \, P_{TW}(t)
\end{equation}
Taking the derivative of this relation and subtituting here the result, Eq.(\ref{c1}), we find
\begin{equation}
 \label{c22}
P_{TW}(s) \; = \; 
\exp\biggl[
-\int_{s}^{\infty} dt \, (t-s) q^{2}(t) 
\biggr] \times \int_{s}^{\infty} dt \, q^{2}(t)
\end{equation}

In the limit $s\to\infty$ the function $q(s)$, according to its definition, eq.(\ref{c11}),
must go to zero, and in this case Eq.(\ref{c20})
turns into the Airy function equation, $q'' = s q$. Thus
\begin{equation}
 \label{c23}
q(s\to\infty) \; \simeq \; \Ai(s) \; \sim \; \exp\Bigl[-\frac{2}{3} s^{3/2}\Bigr]
\end{equation}
It can be proved \cite{Hastings} that in the opposite limit, $s\to -\infty$, the asymptotic form 
of the solution of the Panleve\'e equation (\ref{c20}) (which has the right tail Airy function
limit, Eq.(\ref{c23})) is
\begin{equation}
 \label{c24}
q(s\to -\infty) \; \simeq \; \sqrt{-\frac{1}{2}s}
\end{equation}
Substituting the above two asymptotics into Eq.(\ref{c22}), we can estimate the asymptotic behaviour
for the right and the left tails of the TW probability distribution function:
\begin{eqnarray}
 \label{c25}
P_{TW}(s\to +\infty) &\sim&
 \exp\Bigl[-\frac{4}{3} s^{3/2}\Bigr]
\\
\nonumber
\\
\label{c26}
P_{TW}(s\to -\infty) &\sim&
 \exp\Bigl[-\frac{1}{12} |s|^{3}\Bigr]
\end{eqnarray}

\newpage


\end{document}